\documentstyle{article}    

\def\PPh#1{\setbox0\hbox{$#1\rm I$}\mathord{\vcenter{\ialign{$#1\rm##$\cr
I\cr\noalign{\nointerlineskip \vskip-0.541\ht0}P\cr}}}}

\def\Ph{{\mathpalette\PPh{}}}
\def\Ed{{\mathord{\mkern5mu\mathaccent"7020{\mkern-5mu\partial}}}}

\newcommand{\Edt}{\tilde{\Ed}}
\newcommand{\Pht}{\tilde{\Ph}}

\let\Eqnarray=\eqnarray
\renewcommand{\eqnarray}{\arraycolsep=0.1675em \Eqnarray}

\newcommand{\art}[6]{{\rm #1, \rm #2, \it #3 \bf #4 \rm (#5), \mbox{#6}.}}

\newcommand{\book}[3]{{\rm #1, \it #2, \rm #3.}}

\newtheorem{proposition}{Proposition}[section]
\newtheorem{theorem}[proposition]{Theorem}

\newtheorem{lemma}[proposition]{Lemma}
\newtheorem{definition}[proposition]{Definition}
\newtheorem{example}[proposition]{Example}
\newenvironment{proof}{\noindent{\bf Proof:$\;$}}{\hfill$\Box$\medskip}

\let\Definition=\definition
\renewcommand{\definition}{\Definition \rm}
\let\Example=\example
\renewcommand{\example}{\Example \rm}

\newcommand{\eps}{\varepsilon}
\newcommand{\ch}{{\raise 0.69mm \hbox{$\chi$}}}

\begin{document}

\centerline{\Large{\bf   LANCZOS POTENTIALS AND CURVATURE-
}}

\centerline{\Large{\bf   FREE CONNECTIONS ALIGNED TO A 
}}

\centerline{\Large{\bf   GEODESIC SHEAR-FREE EXPANDING 
}}

\centerline{\Large{\bf   NULL CONGRUENCE
}}

\

\centerline{{\bf Fredrik Andersson}}

\centerline{ \it Department of Mathematics,
 Link\"oping University,}

\centerline{\it S-581 83 Link\"oping,}

\centerline{\it Sweden.}

\centerline{Email: frand@mai.liu.se}

\

\centerline{{\bf Abstract}}

\

By the method of $\rho$-integration we obtain all Lanczos potentials 
$L_{ABCA'}$ of the Weyl spinor that, in a certain sense, are aligned to
a geodesic shear-free expanding null congruence. We also obtain all
spinors $H_{ABA'B'}=Q_{AB}o_{A'}o_{B'}$, $Q_{AB}=Q_{(AB)}$ satisfying 
$\nabla_{(A}{}^{B'}H_{BC)A'B'}=L_{ABCA'}$. We go on to prove that 
$H_{ABA'B'}$ can be chosen so that $\Gamma_{ABCA'}=\nabla_{(A}{}^{B'}
H_{B)CA'B'}$ defines a metric asymmetric curvature-free connection 
such that $L_{ABCA'}=\Gamma_{(ABC)A'}$ is a Lanczos potential that is
aligned to the geodesic shear-free expanding congruence. These 
results are a generalization to a large class of algebraically 
special spacetimes (including all vacuum ones for which the principal 
null direction is expanding) of the curvature-free connection of the
Kerr spacetime found by Bergqvist and Ludvigsen, which was used in a
construction of quasi-local momentum. In conclusion we give a 
corresponding definition of quasi-local momentum in this more general 
class of spacetimes and examine some of its properties in the special 
case of a Kerr-Schild spacetime.

\


\newpage

\section{Preliminaries}

\subsection{Introduction and conventions}

The purpose of this paper is to determine Lanczos potentials for the Weyl
spinor $\Psi_{ABCD}$ and their $H$-potentials that have a particularly simple
algebraic structure, in a class of spacetimes admitting a geodesic shear-free
expanding null congruence (including all vacuum ones), and to use these
potentials to construct curvature-free asymmetric connections. Such a
construction has already been performed in the Kerr spacetime \cite{BL1},
where the curvature-free connection was used to construct a quasi-local
momentum for the Kerr spacetime. In this section we will give the preliminary
results that we need, concerning Lanczos potentials, $H$-potentials and
$\rho$-integration. In the final part of this section we give an outline of
the remainder of the paper.

We will use spacetime definitions and conventions from \cite{PR1}. In 
particular this means that the metric $g_{ab}$ is assumed to have 
signature $(+--\,\,-)$. We will use spinors for our calculations, but
as all results are local in nature there is no need to postulate the 
existence of a global spinor structure on spacetime. Penrose's abstract index
notation \cite{PR1} will be used throughout this paper; Latin letters will
denote tensor indices, primed and unprimed capital Latin letters will denote
spinor indices. However, on differential forms (completely antisymmetric
tensors) occurring under an integral sign the indices will be suppressed and
the differential form will be written as a bold-faced letter. All spinor 
dyads $(o^{A},\iota^{A})$ will be assumed to be normalized, i.e., 
$o_{A}\iota^{A}=1$. $\nabla_{AA'}$ denotes the Levi-Civita connection, i.e.,
the uniquely defined metric and torsion-free (symmetric) connection on
spacetime.

\subsection{Lanczos potentials and $H$-potentials}

It is well-known \cite{Lanczos}, \cite{BC}, \cite{Illge} that there 
always exists a completely symmetric spinor $L_{ABCA'}=L_{(ABC)A'}$ 
such that
\begin{equation}
	\Psi_{ABCD}=2\nabla_{(A}{}^{A'}L_{BCD)A'}
	\label{}
\end{equation}
where $\Psi_{ABCD}$ is the Weyl spinor. This equation is called the 
Weyl-Lanczos equation and $L_{ABCA'}$ is called a Lanczos potential 
of $\Psi_{ABCD}$. In fact \cite{Illge}, given {\em any} symmetric spinor 
$W_{ABCD}$ it can be shown that it has a Lanczos potential $L_{ABCA'}$.

It is important to note that a Lanczos potential is far from unique. 
It is shown in \cite{Illge} that given any symmetric spinors $W_{ABCD}$,
$\zeta_{BC}$ there exists a Lanczos potential of $W_{ABCD}$ (unique up to 
its values on a spacelike past-compact hypersurface) such that
$$
 \nabla^{AA'}L_{ABCA'}=\zeta_{BC}.
$$
For a recent, very simple proof of this fact, see \cite{AE2}.

The spinor $\zeta_{BC}$ is called the differential gauge of $L_{ABCA'}$ 
and when $\zeta_{BC}=0$, i.e.,
$$
 \nabla^{AA'}L_{ABCA'}=0
$$
$L_{ABCA'}$ is said to be in Lanczos differential gauge. Then 
the Weyl-Lanczos equation can be written
$$
 \Psi_{ABCD}=2\nabla_{A}{}^{A'}L_{BCDA'}.
$$
However, in this paper we will {\em not} impose the Lanczos differential 
gauge condition. Instead we prefer $F_{BC}$ to remain arbitrary and indeed
the Lanczos potentials that we find will only satisfy Lanczos differential
gauge in very special circumstances.

We now take this one step further and ask: Given a symmetric spinor
$L_{ABCA'}$, does there exist a spinor $H_{ABA'B'}$ such that
\begin{equation}
	L_{ABCA'}=\nabla_{(A}{}^{B'}H_{BC)A'B'}
	\label{LHspineqn}
\end{equation}
where $H_{ABA'B'}$ is completely symmetric, i.e., $H_{ABA'B'}=
H_{(AB)(A'B')}$? In the case when $L_{ABCA'}$ is a Lanczos potential of
the Weyl spinor, $H_{ABA'B'}$ would then be a gravitational analogue of
the flat space Hertz potential in electromagnetic theory.

Illge gives a partial answer to this question. He shows \cite{Illge}
that if such a potential exists it has to satisfy a restrictive
condition that is algebraic in the $H$-potential. This rules out the
existence of such a Hertz-like potential in general. However, in Einstein
spacetimes the $H$-potential vanishes from this condition and it turns out
to be possible to prove the existence of a completely symmetric $H$-potential
for an arbitrary symmetric $L_{ABCA'}$ in these spacetimes \cite{AE4}.

We remark that in an ${\cal H}$-space  \cite{KLNT} in `complex general
relativity', it is always possible to find a very simple Lanczos potential
of the Weyl spinor, that in turn has a very simple $H$-potential; however
a general result of the nature of the one in \cite{AE2} does not exist, as
far as we know, for these spaces. 

If we remove the requirement of symmetry over the unprimed indices of
$H_{ABA'B'}$, it follows from \cite{Illge} that such a potential exists
in all spacetimes, but in this paper we will only consider completely 
symmetric $H$-potentials so this result is of limited interest to us.

For a lot of our calculations in this paper we will use the 
GHP-formalism. For a normalized spinor dyad $(o^{A},\iota^{A})$ it is 
conventional to define the dyad components of the Lanczos potential, 
the so-called Lanczos scalars, as
\begin{eqnarray}
  L_{0}=L_{ABCA^{\prime}}o^{A}o^{B}o^{C}o^{A^{\prime}} & \quad L_{4}=
  L_{ABCA^{\prime}}o^{A}o^{B}o^{C}\iota^{A^{\prime}} \nonumber \\ L_{1}=
  L_{ABCA^{\prime}}o^{A}o^{B}\iota^{C}o^{A^{\prime}} & \quad L_{5}=
  L_{ABCA^{\prime}}o^{A}o^{B}\iota^{C}\iota^{A^{\prime}} \nonumber \\
  L_{2}=L_{ABCA^{\prime}}o^{A}\iota^{B}\iota^{C}o^{A^{\prime}} & \quad 
  L_{6}=L_{ABCA^{\prime}}o^{A}\iota^{B}\iota^{C}\iota^{A^{\prime}}
  \nonumber \\ L_{3}=L_{ABCA^{\prime}}\iota^{A}\iota^{B}
  \iota^{C}o^{A^{\prime}} & \quad L_{7}=L_{ABCA^{\prime}}
  \iota^{A}\iota^{B}\iota^{C}\iota^{A^{\prime}}. \label{Lscalars}
\end{eqnarray}
The Weyl-Lanczos equation can then be translated into GHP-formalism:
\begin{eqnarray}
 \frac{1}{2}\Psi_{0} & = & \Ed L_{0}-\Ph L_{4}-\bar{\tau}^{\prime}L_{0}
	+3\sigma L_{1}+\bar{\rho}L_{4}-3\kappa L_{5} \nonumber \\ 2\Psi_{1}
	& = & 3\Ed L_{1}-3\Ph L_{5}-\Ed^{\prime}L_{4}+\Ph^{\prime}L_{0}-(
	\bar{\rho}^{\prime}-3\rho^{\prime})L_{0}-3(\bar{\tau}^{\prime}-\tau)
	L_{1} \nonumber \\ & & +6\sigma L_{2}-(3\tau^{\prime}-\bar{\tau})L_{4}-
	3(\rho-\bar{\rho})L_{5}-6\kappa L_{6} \nonumber \\Ê\Psi_{2} & = &
	\Ed L_{2}-\Ph L_{6}-\Ed^{\prime}L_{5}+\Ph^{\prime}L_{1}+\kappa^{\prime}L_{0}
	-(\bar{\rho}^{\prime}-2\rho^{\prime})L_{1}-(\bar{\tau}^{\prime}-2\tau)
	L_{2} \nonumber	\\ & & +\sigma L_{3}-\sigma^{\prime}L_{4}-(2\tau^{\prime}
	-\bar{\tau})L_{5}-(2\rho-\bar{\rho})L_{6}-\kappa L_{7} \nonumber \\
	2\Psi_{3} & = & \Ed L_{3}-\Ph L_{7}-3\Ed^{\prime}L_{6}+3\Ph^{\prime}L_{2}
	+6\kappa^{\prime}L_{1}-3(\bar{\rho}^{\prime}-\rho^{\prime})L_{2}-(
	\bar{\tau}^{\prime}-3\tau)L_{3} \nonumber \\Ê& & -6\sigma^{\prime}
	L_{5}-3(\tau^{\prime}-\bar{\tau})L_{6}-(3\rho-\bar{\rho})L_{7}
	\nonumber \\Ê\frac{1}{2}\Psi_{4} & = & \Ph^{\prime}L_{3}-\Ed^{\prime}L_{7}
	+3\kappa^{\prime}L_{2}-\bar{\rho}^{\prime}L_{3}-3\sigma^{\prime}L_{6}+
	\bar{\tau}L_{7}
\end{eqnarray}
These equations will be used to integrate the Weyl-Lanczos equation 
for a large class of algebraically special spacetimes in the following
sections.

We define the dyad components of $H_{ABA'B'}$ as
\begin{eqnarray}
 H_{00'}=H_{ABA'B'}o^{A}o^{B}o^{A'}o^{B'} & \quad H_{01'}=H_{ABA'B'}o^{A}o^{B}
 o^{A'}\iota^{B'} \nonumber \\ H_{02'}=H_{ABA'B'}o^{A}o^{B}\iota^{A'}\iota^{B'}
 & \quad H_{10'}=H_{ABA'B'}o^{A}\iota^{B}o^{A'}o^{B'} \nonumber \\ÊH_{11'}=
 H_{ABA'B'}o^{A}\iota^{B}o^{A'}\iota^{B'} & \quad H_{12'}=H_{ABA'B'}o^{A}
 \iota^{B}\iota^{A'}\iota^{B'}\nonumber \\ÊH_{20'}=H_{ABA'B'}\iota^{A}\iota^{B}
 o^{A'}o^{B'} & \quad H_{21'}=H_{ABA'B'}\iota^{A}\iota^{B}o^{A'}\iota^{B'}
 \nonumber \\ H_{22'}=H_{ABA'B'}\iota^{A}\iota^{B}\iota^{A'}\iota^{B'}.
\end{eqnarray}
Then (\ref{LHspineqn}) becomes, in GHP-formalism
\begin{eqnarray}
 L_{0} & = & \Ed H_{00'}-\Ph H_{01'}-\bar{\tau}'H_{00'}+2\bar{\rho}H_{01'}
 -\bar{\kappa}H_{02'}+2\sigma H_{10'}-2\kappa H_{11'} \nonumber \\ 3L_{1}
 & = & \Ph'H_{00'}-\Ed'H_{01'}+2\Ed H_{10'}-2\Ph H_{11'} \nonumber \\ & &
 +(2\rho'-\bar{\rho}')H_{00'}+2(\bar{\tau}-\tau')H_{01'}-\bar{\sigma}H_{02'}
 +2(\tau-\bar{\tau}')H_{10'} \nonumber \\ & & -2(\rho-2\bar{\rho})H_{11'}-2
 \bar{\kappa}H_{12'}+2\sigma H_{20'}-2\kappa H_{21'} \nonumber \\ 3L_{2}
 & = & 2\Ph'H_{10'}-2\Ed'H_{11'}+\Ed H_{20'}-\Ph H_{21'} \nonumber \\ & &
 +2\kappa'H_{00'}-2\sigma'H_{01'}+2(\rho'-\bar{\rho}')H_{10'}+2(2\bar{\tau}
 -\tau')H_{11'} \nonumber \\ & & -2\bar{\sigma}H_{12'}+(2\tau-\bar{\tau}')
 H_{20'}-2(\rho-\bar{\rho})H_{21'}-\bar{\kappa}H_{22'} \nonumber \\ L_{3}
 & = & \Ph'H_{20'}-\Ed'H_{21'}+2\kappa'H_{10'}-2\sigma'H_{11'}-\bar{\rho}'
 H_{20'}+2\bar{\tau}H_{21'}-\bar{\sigma}H_{22'} \nonumber \\ L_{4} & = &
 \Ed H_{01'}-\Ph H_{02'}+\bar{\sigma}'H_{00'}-2\bar{\tau}'H_{01'}+\bar{\rho}
 H_{02'}+2\sigma H_{11'}-2\kappa H_{12'} \nonumber \\ 3L_{5} & = & \Ph'H_{01'}
 -\Ed'H_{02'}+2\Ed H_{11'}-2\Ph H_{12'} \nonumber \\Ê& & +\bar{\kappa}'H_{00'}
 +2(\rho'-\bar{\rho}')H_{01'}+(\bar{\tau}-2\tau')H_{02'}+2\bar{\sigma}'H_{10'}
 \nonumber \\ & & +2(\tau-2\bar{\tau}')H_{11'}-2(\rho-\bar{\rho})H_{12'}+2
 \sigma H_{21'}-2\kappa H_{22'} \nonumber \\ 3L_{6} & = & 2\Ph'H_{11'}-2\Ed'
 H_{12'}+\Ed H_{21'}-\Ph H_{22'} \nonumber \\ & & +2\kappa'H_{01'}-2\sigma'
 H_{02'}+2\bar{\kappa}'H_{10'}+2(\rho'-2\bar{\rho}')H_{11'} \nonumber \\ & &
 +2(\bar{\tau}-\tau')H_{12'}+\bar{\sigma}'H_{20'}+2(\tau-\bar{\tau}')H_{21'}
 -(2\rho-\bar{\rho})H_{22'} \nonumber \\ L_{7} & = & \Ph'H_{21'}-\Ed'H_{22'}
 +2\kappa'H_{11'}-2\sigma'H_{12'}+\bar{\kappa}'H_{20'}-2\bar{\rho}'H_{21'}
 +\bar{\tau}H_{22'}
 \label{LHGHPeqn}
\end{eqnarray}
and we will also integrate these equations for the Lanczos potentials 
obtained from the GHP Weyl-Lanczos equations.

\subsection{Some spacetimes admitting a geodesic shear-free expanding null 
congruence}

In \cite{Andersson2} the GHP-equations for the spin coefficients and 
curvature components were $\rho$-integrated. In this section we will
simply quote the results. We assume that spacetime admits a geodesic,
shear-free null congruence $l^{a}=o^{A}o^{A'}$ and that its Ricci spinor
satisfies the condition
\begin{equation}
 \Phi_{ABA'B'}o^{A}o^{B}=0.
\end{equation}
For various technical reasons we also restrict the scalar curvature
to be constant and the null congruence to be expanding. Any spacetime that
satisfies all these conditions will be said to be of class ${\cal G}$. Take
$o^{A}$ as the first spinor of a spinor dyad. In GHP-formalism the above
conditions are equivalent to
\begin{equation}
 \Phi_{00}=\Phi_{01}=\Phi_{02}=0\quad,\quad\kappa=\sigma=0\quad,\quad
 \rho\neq 0\quad,\quad\Lambda={\rm constant}
\end{equation}
By the Goldberg-Sachs theorem we obtain
\begin{equation}
 \Psi_{0}=\Psi_{1}=0,
\end{equation}
and so the spacetime is algebraically special.

We can use a null rotation about $o^{A}$ to achieve $\tau=0$, and the Ricci
equations \cite{PR1} then imply that also
\begin{equation}
 \tau'=\sigma'=0.
\end{equation}
Whenever a dyad is chosen in this way for an arbitrary spacetime of 
class ${\cal G}$ it will be said to be in standard form.

We introduce Held's \cite{Held} modified operators which can be written
\begin{equation}
 \Edt=\frac{1}{\bar{\rho}}\Ed \quad ,\quad\Edt'=\frac{1}{\rho}\Ed'
 \quad ,\quad \Pht'=\Ph'+\frac{p}{2\rho}(\Psi_{2}+2\Lambda)+
 \frac{q}{2\bar{\rho}}(\bar{\Psi}_{2}+2\Lambda)
\end{equation}
in this dyad. Note that our definition of $\Pht'$ is slightly modified 
from Held's (by the inclusion of $\Lambda$ in the non-vacuum case). The
purpose of using Held's modified operators is simply to reduce the length
of calculations; in particular the new operators have the nice properties
\begin{equation}
 \Bigl[\Ph,\Edt\Bigr]=\Bigl[\Ph,\Edt'\Bigr]=0
\end{equation}
and
\begin{equation}
	\Bigl[\Ph,\Pht'\Bigr]\eta=\bigl[-\frac{1}{2\rho}(\Psi_{2}+2\Lambda)-
	\frac{1}{2\bar{\rho}}(\bar{\Psi}_{2}+2\Lambda)\bigr]\Ph\eta
	\label{}
\end{equation}
so that, in particular, if $\eta^{\circ}$ satisfies $\Ph \eta^{\circ}=0$
(a degree sign will throughout the paper, be used to denote a quantity that
is killed by $\Ph$) then
$$
\Ph \Edt'\eta^{\circ}=\Bigl[\Ph,\Edt'\Bigr] \eta^{\circ}=0
$$
and the same result is true if $\Edt'$ is replaced with $\Edt$ or $\Pht'$.

We will now give the results of the integration. More details can be 
found in \cite{Andersson2}.

First of all, the GHP-operators acting on $\rho$ are
\begin{eqnarray}
 \Ph \rho & = & \rho^2 \nonumber \\ \Edt \rho & = & 0 \nonumber \\ \Edt'
 \rho & = & \rho^2 \Edt'\Omega^{\circ} \nonumber \\ \Pht'\rho & = & \rho^2
 \bar{\rho}'^{\circ}-\frac{1}{2}\rho^2\bar{\rho}\overline{\Psi}_{2}^{\circ}
 -\frac{1}{2}\rho^3\Psi_{2}^{\circ}-\rho^3\bar{\rho}\Phi_{11}^{\circ}+
 \frac{\rho}{\bar{\rho}}\Lambda
\end{eqnarray}
where $\Omega^{\circ}=\frac{1}{\bar{\rho}}-\frac{1}{\rho}$ is the 
twist of the congruence. From these we obtain the useful relations
\begin{eqnarray}
 \Ph\Omega^{\circ} & = & 0 \nonumber \\ \Pht'\Omega^{\circ} & = &
 \bar{\rho}'^{\circ}-\rho'^{\circ} \nonumber \\ \Edt\Edt'\Omega^{\circ}
 & = & 2\Omega^{\circ}\bar{\rho}'^{\circ}+\Psi_{2}^{\circ}
 -\overline{\Psi}_{2}^{\circ}
\end{eqnarray}
The curvature scalars and the spin coefficients are
\begin{eqnarray}
 \rho' & = & \bar{\rho}\rho'^{\circ}-\frac{1}{2}(\rho^2+\rho\bar{\rho})
 \Psi_{2}^{\circ}-\rho^2\bar{\rho}\Phi_{11}^{\circ}+\frac{1}{\bar{\rho}}
 \Lambda \nonumber \\ \kappa' & = & \kappa'^{\circ}-\rho\Psi_{3}^{\circ}
 -\frac{1}{2}\rho^2\Edt'\Psi_{2}^{\circ}-\frac{1}{2}\rho^3\Psi_{2}^{\circ}
 \Edt'\Omega^{\circ}-\rho\bar{\rho}\Phi_{21}^{\circ}-\rho^2\bar{\rho}
 \Edt'\Phi_{11}^{\circ}-\rho^3\bar{\rho}\Phi_{11}^{\circ}\Edt'
 \Omega^{\circ} \nonumber \\ \Psi_{2} & = & \rho^3\Psi_{2}^{\circ}+2\rho^3
 \bar{\rho}\Phi_{11}^{\circ} \nonumber \\ \Psi_{3} & = & \rho^2
 \Psi_{3}^{\circ}+\rho^3\Edt' \Psi_{2}^{\circ}+\frac{3}{2}\rho^4
 \Psi_{2}^{\circ}\Edt'\Omega^{\circ}+\rho^2\bar{\rho}\Phi_{21}^{\circ}+2
 \rho^3\bar{\rho}\Edt'\Phi_{11}^{\circ}+3\rho^4\bar{\rho}\Phi_{11}^{\circ}
 \Edt' \Omega^{\circ} \nonumber \\ \Psi_{4} & = & \rho\Psi_{4}^{\circ}+\rho^2
 \Edt'\Psi_{3}^{\circ}+\frac{1}{2}\rho^3\bigl(\Edt'^2\Psi_{2}^{\circ}+2
 \Psi_{3}^{\circ}\Edt'\Omega^{\circ}\bigr)+\frac{1}{2}\rho^4\bigl(
 \Psi_{2}^{\circ}\Edt'^2\Omega^{\circ}+3\Edt'\Omega^{\circ}\Edt'
 \Psi_{2}^{\circ}\bigr) \nonumber \\ & & +\frac{3}{2}\rho^5\Psi_{2}^{\circ}
 (\Edt'\Omega^{\circ})^2+\rho^2\bar{\rho}\Edt'\Phi_{21}^{\circ}+\rho^3
 \bar{\rho}\bigl(\Edt'^2\Phi_{11}^{\circ}+\Phi_{21}^{\circ}\Edt'
 \Omega^{\circ}\bigr) \nonumber \\ & & +\rho^4\bar{\rho}\bigl(
 \Phi_{11}^{\circ}\Edt'^2\Omega^{\circ}+3\Edt'\Omega^{\circ}\Edt'
 \Phi_{11}^{\circ}\bigr)+3\rho^5\bar{\rho}\Phi_{11}^{\circ}(\Edt'
 \Omega^{\circ})^2 \nonumber \\ \Phi_{11} & = & \rho^2\bar{\rho}^2
 \Phi_{11}^{\circ} \nonumber \\ \Phi_{21} & = & \rho\bar{\rho}^2
 \Phi_{21}^{\circ}+\rho^2\bar{\rho}^2 \Edt' \Phi_{11}^{\circ}+\rho^3
 \bar{\rho}^2\Phi_{11}^{\circ}\Edt'\Omega^{\circ} \nonumber \\ 
 \Phi_{22} & = & \rho\bar{\rho}\Phi_{22}^{\circ}+\rho^2\bar{\rho}\bigl(\Edt'
 \overline{\Phi}_{21}^{\circ}-\frac{1}{2}\Pht'\Phi_{11}^{\circ}\bigr)
 +\rho\bar{\rho}^2\bigl(\Edt\Phi_{21}^{\circ}-\frac{1}{2}\Pht'\Phi_{11}^{\circ}
 \bigr) \nonumber \\Ê& & +\rho^3\bar{\rho}\overline{\Phi}_{21}^{\circ}\Edt'
 \Omega^{\circ}+\frac{1}{2}\bigl(\Edt\Edt'\Phi_{11}^{\circ}+\Edt'\Edt
 \Phi_{11}^{\circ}\bigr)-\rho\bar{\rho}^3\Phi_{21}^{\circ}\Edt\Omega^{\circ}
 \nonumber \\ & & +\rho^3\bar{\rho}^2 \Edt'\Omega^{\circ}\Edt\Phi_{11}^{\circ}
 -\rho^2\bar{\rho}^3\Edt\Omega^{\circ}\Edt'\Phi_{11}^{\circ}-\rho^3
 \bar{\rho}^3\Phi_{11}^{\circ}\Edt\Omega^{\circ}\Edt'\Omega^{\circ}.
\end{eqnarray}
The remaining Ricci and Bianchi equations are
\begin{eqnarray}
 \Pht'\rho'^{\circ}-\Edt\kappa'^{\circ} & = & \Lambda(2\Omega^{\circ}
 \rho'^{\circ}+\Psi_{2}^{\circ}-\overline{\Psi}_{2}^{\circ}) \nonumber \\
 \Edt'\kappa'^{\circ} & = & -\Psi_{4}^{\circ} \nonumber \\ \Edt'
 \rho'^{\circ} & = & -\Omega^{\circ}\kappa'^{\circ}-\Psi_{3}^{\circ}
 \nonumber \\ \Edt\Psi_{2}^{\circ} & = & 2\overline{\Phi}_{21}^{\circ}
 \nonumber \\ \Edt\Psi_{3}^{\circ}-\Pht'\Psi_{2}^{\circ} & = &
 \Phi_{22}^{\circ} \nonumber \\ \Edt\Psi_{4}^{\circ}-\Pht'\Psi_{3}^{\circ}
 & = & \Lambda\bigl(\Edt'\Psi_{2}^{\circ}-2\Phi_{21}^{\circ}-2\Omega^{\circ}
 \Psi_{3}^{\circ}\bigr). \label{RicBian}
\end{eqnarray}
Finally, the commutators become
\begin{eqnarray}
 \Bigl[\Ph,\Edt\Bigr] & = & 0 \nonumber \\ \Bigl[\Ph,\Edt'\Bigr] & = & 0
 \nonumber \\Ê\Bigl[\Ph,\Pht'\Bigr] & = & -\Bigl(\frac{1}{2}\rho^2
 \Psi_{2}^{\circ}+\frac{1}{2}\bar{\rho}^2\overline{\Psi}_{2}^{\circ}+
 \rho^2\bar{\rho}\Phi_{11}^{\circ}+\rho\bar{\rho}^2\Phi_{11}^{\circ}+
 \Lambda\bigl(\frac{1}{\rho}+\frac{1}{\bar{\rho}}\bigr)\Bigr)\Ph
 \nonumber \\ \Bigl[\Pht',\Edt\Bigr] & = & \Bigl(-\frac{\bar{\kappa}'^{\circ}}
 {\bar{\rho}}+\overline{\Psi}_{3}^{\circ}+\frac{1}{2}\bar{\rho}\Edt
 \overline{\Psi}_{2}^{\circ}-\frac{1}{2}\bar{\rho}^2\overline{\Psi}_{2}^{\circ}
 \Edt\Omega^{\circ}+\rho\overline{\Phi}_{21}^{\circ}+\rho\bar{\rho}\Edt
 \Phi_{11}^{\circ} \nonumber \\Ê& & -\rho\bar{\rho}^2\Phi_{11}^{\circ}\Edt
 \Omega^{\circ}\Bigr)\Ph+q\bigl(\bar{\kappa}'^{\circ}-\Lambda\Edt\Omega^{\circ}
 \bigr)\nonumber \\ \Bigl[\Pht',\Edt'\Bigr] & = & \Bigl(-\frac{\kappa'^{\circ}}
 {\rho}+\Psi_{3}^{\circ}+\frac{1}{2}\rho\Edt'\Psi_{2}^{\circ}+\frac{1}{2}\rho^2
 \Psi_{2}^{\circ}\Edt'\Omega^{\circ}+\bar{\rho}\Phi_{21}^{\circ}+\rho
 \bar{\rho}\Edt'\Phi_{11}^{\circ} \nonumber \\ & & +\rho^2\bar{\rho}
 \Phi_{11}^{\circ}\Edt'\Omega^{\circ}\Bigr)\Ph+p\bigl(\kappa'^{\circ}+
 \Lambda\Edt'\Omega^{\circ}\bigr) \nonumber \\ \Bigl[\Edt,\Edt'\Bigr] & = &
 \Bigl(\frac{\bar{\rho}'^{\circ}}{\bar{\rho}}-\frac{\rho'^{\circ}}{\rho}+
 \frac{\rho}{2}\bigl(\frac{1}{\rho}+\frac{1}{\bar{\rho}}\bigr)\Psi_{2}^{\circ}
 -\frac{\bar{\rho}}{2}\bigl(\frac{1}{\rho}+\frac{1}{\bar{\rho}}\bigr)
 \overline{\Psi}_{2}^{\circ}+\Omega^{\circ}(\rho\bar{\rho}\Phi_{11}^{\circ}
 -\Lambda)\Bigr)\Ph \nonumber \\Ê& & +\Omega^{\circ}\Pht'+p\bigl(\rho'^{\circ}
 +\Omega^{\circ 2}\Lambda\bigr)-q(\bar{\rho}'^{\circ}+\Omega^{\circ 2}
 \Lambda\bigr)
\end{eqnarray}
It is worth noting that the sixth equation of (\ref{RicBian}) and the 
imaginary part of the fifth equation of (\ref{RicBian}) are actually 
consequences of the other equations.

\subsection{Outline}

In Section 2 we $\rho$-integrate the Weyl-Lanczos equations and obtain
their general solution in the case when $L_{ABCA'}=M_{ABC}o_{A'}$, 
for spacetimes of class ${\cal G}$ where $l^{a}=o^{A}o^{A'}$ is the
geodesic shear-free expanding null-congruence.

In Section 3 we consider the equation
$$
 L_{ABCA'}=\nabla_{(A}{}^{B'}H_{BC)A'B'}
$$
where $L_{ABCA'}$ is found in Section 2 and $H_{ABA'B'}$ is completely
symmetric; we use the results and techniques from Section 2 to find its
general solution for the case $H_{ABA'B'}=Q_{AB}o_{A'}o_{B'}$. In particular
it is shown that such an $H$-potential always exists, providing the 
function of integration $L_{7}^{\circ}$ from Section 2 vanishes, which 
is a permissible choice.

Section 4 concerns itself with metric connections $\hat{\nabla}_{AA'}$
defined by
\begin{equation}
 \hat{\nabla}_{AA'}\xi^{B}=\nabla_{AA'}\xi^{B}+2\Gamma_{C}{}^{B}{}_{AA'}
 \xi^{C} \label{Conndef}
\end{equation}
where
$$
 \Gamma_{ABCA'}=L_{ABCA'}+\eps_{AC}\ch_{BA'}+\eps_{BC}\ch_{AA'}
$$
and $L_{ABCA'}$ is symmetric over its unprimed indices. We remark 
that a spacetime equipped with such a connection is called a 
Riemann-Cartan spacetime. It has been shown \cite{BL1} that {\em in the
Kerr spacetime} a particular choice of such a connection, due to the fact
that it has vanishing curvature, can be used to define quasi-local
momentum. This particular choice of $\Gamma_{ABCA'}$ can also be written
$$
 \Gamma_{ABCA'}=\nabla_{(A}{}^{B'}H_{B)CA'B'},
$$
where $H_{ABA'B'}=Q_{AB}o_{A'}o_{B'}$ for some spinor $Q_{AB}=Q_{(AB)}$.

It was subsequently shown \cite{Bergqvist} for this choice of 
$\Gamma_{ABCA'}$, that the symmetric part $L_{ABCA'}$ is actually a Lanczos 
potential of the Weyl spinor in the Kerr spacetime. It is therefore of 
interest to see if the Lanczos- and $H$-potentials found in Section 3 
and 4 can be used to define a connection that has vanishing curvature
for these more general spacetimes.

We show that any connection $\hat{\nabla}_{AA'}$ defined by (\ref{Conndef})
from a Lanczos potential of the type investigated in Section 2, has vanishing
Weyl curvature, i.e., $\hat{\Psi}_{ABCD}=0$. We also show that we can
accomplish $\hat{\Sigma}_{AB}=0$ if and only if the Lanczos potential we
start from possesses an $H$-potential of the type investigated in Section 3.
We go on to prove that in spacetimes where $\Lambda=0$ or
$\Edt'\Omega^{\circ}=0$ we can also eliminate $\hat{\Lambda}$ by choosing
the functions of integration $L_{6}^{\circ}=-\Lambda$ and $H_{22'}^{\circ}=
-\frac{3}{2}\Edt' L_{5}^{\circ}-\Omega^{\circ}\Lambda$.

When we look at the Ricci spinor $\hat{\Phi}_{ABA'B'}$ it is shown 
that three of its components always vanish, and providing $\Lambda=0$
the remaining six components can be eliminated by fixing another function
of integration $H_{12'}^{\circ}=3\Omega^{\circ}L_{5}^{\circ}$ and
demanding that the three remaining functions of integration
$L_{4}^{\circ}$, $L_{5}^{\circ}$ and $H_{02'}^{\circ}$ are solutions of a
coupled system of third order equations involving only the differential
operator $\Edt'$, and a first order non-linear equation involving the
operators $\Edt$ and $\Pht'$ only. We go on to prove that all these conditions
can be simultaneously satisfied and hence, providing $\Lambda=0$, a completely
curvature-free connection can always be constructed in this manner.

In Section 5 we examine the Bergqvist-Ludvigsen construction of 
quasi-local momentum in class ${\cal G}$ spacetimes with vanishing 
Ricci scalar, and in greater detail in the special case of 
Kerr-Schild spacetimes belonging to this class.

Section 6 discusses possible ways of continuing this work, and also 
contains a few concluding remarks. 

\section{All Lanczos potentials of the Weyl spinor that are aligned to 
$o^{A'}$}

In this section we will find all Lanczos potentials of $\Psi_{ABCD}$ 
in spacetimes of class ${\cal G}$, that have the algebraic structure
$L_{ABCA'}=M_{ABC}o_{A'}$ with $o^{A}$ as in the previous section. Such a 
Lanczos potential will be said to be aligned to $o^{A'}$. Thus, we 
assume once again that we have a spacetime of class ${\cal G}$ with a 
spinor dyad in standard form. That $L_{ABCA'}$ is aligned to $o^{A'}$
amounts to choosing the Lanczos scalars
\begin{equation}
	L_{0}=L_{1}=L_{2}=L_{3}=0
	\label{}
\end{equation}
The existence of such Lanczos potentials in these spacetimes has 
already been shown by Torres del Castillo \cite{TdC1}, \cite{TdC2}. He 
actually proves existence in the slightly more general class of 
spacetimes that does not require $\Lambda$ to be a constant and also 
allows $\rho=0$. However, his approach differs significantly from
ours and he is therefore unable to find {\em all} Lanczos potentials
of this type. 

The Weyl-Lanczos equations in GHP-formalism then become
\begin{eqnarray}
 0 & = & -\Ph L_{4}+\bar{\rho}L_{4} \nonumber \\ 0 & = & -3\Ph L_{5}
 -\rho\Edt' L_{4}-3(\rho-\bar{\rho})L_{5} \nonumber \\ \Psi_{2} & = &
 -\Ph L_{6}-\rho\Edt' L_{5}-(2\rho-\bar{\rho})L_{6} \nonumber 
 \\Ê2\Psi_{3} & = & -\Ph L_{7}-3\rho\Edt' L_{6}-(3\rho-\bar{\rho})L_{7}
 \nonumber \\Ê\frac{1}{2}\Psi_{4} & = & -\rho\Edt' L_{7}.
 \label{GHPWL} 
\end{eqnarray}
The first equation can immediately be $\rho$-integrated:
$$
 0=\frac{1}{\bar{\rho}}\Ph L_{4}-L_{4}=\Ph\Bigl(\frac{L_{4}}{\bar{\rho}}
 \Bigr)
$$
so that
\begin{equation}
	L_{4}=\bar{\rho}L_{4}^{\circ}.
	\label{}
\end{equation}
Then
$$
 \Edt' L_{4}=\bar{\rho}\Edt' L_{4}^{\circ}
$$
which substituted into the second equation gives
$$
 0=\frac{\rho}{\bar{\rho}}\Ph L_{5}+\Bigl(\frac{\rho^2}{\bar{\rho}}-
 \rho\Bigr)L_{5}+\frac{1}{3}\rho^2\Edt' L_{4}^{\circ}=\Ph\Bigl(
 \frac{\rho}{\bar{\rho}}L_{5}+\frac{1}{3}\rho\Edt' L_{4}^{\circ}\Bigr)
$$
Thus,
\begin{equation}
	L_{5}=\frac{\bar{\rho}}{\rho}L_{5}^{\circ}-\frac{1}{3}\bar{\rho}
	\Edt' L_{4}^{\circ}
	\label{}
\end{equation}
Substituting this into the third equation and $\rho$-integrating in 
the same way gives, using the expression for $\Psi_{2}$, an expression
for $L_{6}$
\begin{equation}
	L_{6}=\frac{\bar{\rho}}{\rho^2}L_{6}^{\circ}-\frac{\bar{\rho}}{\rho}
	\Edt' L_{5}^{\circ}+\frac{1}{6}\bar{\rho}\bigl(\Edt'^2 L_{4}^{\circ}
	+3L_{5}^{\circ}\Edt' \Omega^{\circ}\bigr)-\frac{1}{4}\rho^2
	\Psi_{2}^{\circ}-\frac{1}{12}\rho\bar{\rho}\Psi_{2}^{\circ}-
	\frac{1}{2}\rho^2\bar{\rho}\Phi_{11}^{\circ}.
	\label{}
\end{equation}
We can also $\rho$-integrate the fourth equation to get an expression
for $L_{7}$
\begin{eqnarray}
 L_{7} & = & \frac{\bar{\rho}}{\rho^3}L_{7}^{\circ}-3\frac{\bar{\rho}}
 {\rho^2}\Edt'L_{6}^{\circ}+\frac{3}{2}\frac{\bar{\rho}}{\rho}\bigl(
 \Edt'^2L_{5}^{\circ}+2L_{6}^{\circ}\Edt'\Omega^{\circ}\bigr)-\frac{1}{2}
 \rho\Psi_{3}^{\circ} \nonumber \\ & & -\frac{1}{6}\bar{\rho}\bigl(\Edt'^3
 L_{4}^{\circ}+3L_{5}^{\circ}\Edt'^2 \Omega^{\circ}+9\Edt' \Omega^{\circ}
 \Edt' L_{5}^{\circ}+\Psi_{3}^{\circ}\bigr)-\frac{1}{4}\rho^2\Edt'
 \Psi_{2}^{\circ}-\frac{1}{2}\rho\bar{\rho}\Phi_{21}^{\circ} \nonumber 
 \\ & & -\frac{1}{4}\rho^3\Psi_{2}^{\circ}\Edt' \Omega^{\circ}-\frac{1}{2}
 \rho^2\bar{\rho}\Edt'\Phi_{11}^{\circ}-\frac{1}{2}\rho^3\bar{\rho}
 \Phi_{11}^{\circ}\Edt' \Omega^{\circ}
\end{eqnarray}
These Lanczos scalars will give a Lanczos potential if and only if 
the fifth equation of (\ref{GHPWL}) is satisfied. By substituting the 
above expression for $L_{7}$ into this equation, and using the formula
$$
 \bar{\rho}=\frac{\rho}{1+\rho\Omega^{\circ}}
$$
we find that the fifth equation of (\ref{GHPWL}) is satisfied if and 
only if
\begin{eqnarray}
 0 & = & \Edt' L_{7}^{\circ}-3\rho\bigl(\Edt'^2 L_{6}^{\circ}+L_{7}^{\circ}
 \Edt' \Omega^{\circ}\bigr)+\frac{3}{2}\rho^2 \bigl(\Edt'^3
 L_{5}^{\circ}+2L_{6}^{\circ}\Edt'^2\Omega^{\circ}+6\Edt' \Omega^{\circ}\Edt'
 L_{6}^{\circ}+\frac{1}{3}\Psi_{4}^{\circ}\bigr) \nonumber \\ & & -\frac{1}{6}
 \rho^3 \bigl(\Edt'^4 L_{4}^{\circ}+3L_{5}^{\circ}\Edt'^3 \Omega^{\circ}+12
 \Edt'^2 \Omega^{\circ}\Edt' L_{5}^{\circ}+18\Edt' \Omega^{\circ}\Edt'^2
 L_{5}^{\circ}+18L_{6}^{\circ}(\Edt' \Omega^{\circ})^2 \nonumber \\ & & +
 \Edt' \Psi_{3}^{\circ}-3\Omega^{\circ}\Psi_{4}^{\circ}\bigr).
 \label{GHPWL5}
\end{eqnarray}
By repeatedly applying $\Ph$ to the RHS of the above expression, and 
dividing by $\rho^2$, it is easy to show that equation (\ref{GHPWL5}) is 
satisfied if and only if each coefficient vanishes. Thus, the above 
Lanczos scalars will yield a Lanczos potential of $\Psi_{ABCD}$ if 
and only if the functions of integration satisfy
\begin{eqnarray}
 0 & = & \Edt' L_{7}^{\circ} \nonumber \\Ê0 & = & \Edt'^2 L_{6}^{\circ}
 +L_{7}^{\circ}\Edt' \Omega^{\circ} \nonumber \\ 0 & = & \Edt'^3
 L_{5}^{\circ}+2L_{6}^{\circ}\Edt'^2 \Omega^{\circ}+6\Edt' \Omega^{\circ}
 \Edt' L_{6}^{\circ}+\frac{1}{3}\Psi_{4}^{\circ} \nonumber \\ 0 & = &
 \Edt'^4 L_{4}^{\circ}+3L_{5}^{\circ}\Edt'^3 \Omega^{\circ}+12\Edt'^2
 \Omega^{\circ}\Edt' L_{5}^{\circ}+18\Edt' \Omega^{\circ}\Edt'^2
 L_{5}^{\circ}+18L_{6}^{\circ}(\Edt' \Omega^{\circ})^2 \nonumber 
 \\ & & +\Edt' \Psi_{3}^{\circ}-3\Omega^{\circ}\Psi_{4}^{\circ}
 \label{LpotConditions}
\end{eqnarray}
Since $\Bigl[\Ph,\Edt'\Bigr]=0$ it follows that the first of the above 
equations can locally be solved for $L_{7}^{\circ}$. Once we have done 
that, the second equation can be solved for $L_{6}^{\circ}$. 
Similarly, the third and fourth equation can be solved for 
$L_{5}^{\circ}$ and $L_{4}^{\circ}$ respectively, irrespective of the 
values of $\Omega^{\circ}$, $\Psi_{3}^{\circ}$ and $\Psi_{4}^{\circ}$.
Hence, we have proved the following theorem:

\begin{theorem}
 For any spacetime of class ${\cal G}$ with spinor dyad in standard 
 form, all Lanczos potentials of the Weyl spinor that are aligned to $o^{A'}$
 are given by
 \begin{eqnarray}
  L_{4} & = & \bar{\rho}L_{4}^{\circ} \nonumber \\ L_{5} & = &
  \frac{\bar{\rho}}{\rho}L_{5}^{\circ}-\frac{1}{3}\bar{\rho}\Edt'
  L_{4}^{\circ} \nonumber \\ L_{6} & = & \frac{\bar{\rho}}{\rho^2}
  L_{6}^{\circ}-\frac{\bar{\rho}}{\rho}\Edt' L_{5}^{\circ}+\frac{1}{6}
  \bar{\rho}\bigl(\Edt'^2 L_{4}^{\circ}+3L_{5}^{\circ}\Edt' \Omega^{\circ}
  \bigr)-\frac{1}{4}\rho^2\Psi_{2}^{\circ}-\frac{1}{12}\rho\bar{\rho}
  \Psi_{2}^{\circ}-\frac{1}{2}\rho^2\bar{\rho}\Phi_{11}^{\circ} 
  \nonumber \\ L_{7} & = & \frac{\bar{\rho}}{\rho^3}L_{7}^{\circ}-3
  \frac{\bar{\rho}}{\rho^2}\Edt'L_{6}^{\circ}+\frac{3}{2}\frac{\bar{\rho}}
  {\rho}\bigl(\Edt'^2L_{5}^{\circ}+2L_{6}^{\circ}\Edt'\Omega^{\circ}\bigr)-
  \frac{1}{2}\rho\Psi_{3}^{\circ} \nonumber \\ & & -\frac{1}{6}\bar{\rho}
  \bigl(\Edt'^3L_{4}^{\circ}+3L_{5}^{\circ}\Edt'^2 \Omega^{\circ}+9\Edt'
  \Omega^{\circ}\Edt' L_{5}^{\circ}+\Psi_{3}^{\circ}\bigr)-\frac{1}{4}\rho^2
  \Edt'\Psi_{2}^{\circ}-\frac{1}{2}\rho\bar{\rho}\Phi_{21}^{\circ} \nonumber 
  \\ & & -\frac{1}{4}\rho^3\Psi_{2}^{\circ}\Edt' \Omega^{\circ}-\frac{1}{2}
  \rho^2\bar{\rho}\Edt'\Phi_{11}^{\circ}-\frac{1}{2}\rho^3\bar{\rho}
  \Phi_{11}^{\circ}\Edt' \Omega^{\circ}
  \label{Lscalars}
 \end{eqnarray}
 where the functions $L_{4}^{\circ}, L_{5}^{\circ}, L_{6}^{\circ}$ and
 $L_{7}^{\circ}$ are subject to the conditions (\ref{LpotConditions}).
 In particular, there always exists a local Lanczos potential that is 
 aligned to $o^{A'}$.
\end{theorem}
For future reference, we note that a particular solution of the first two 
equations (\ref{LpotConditions}) is $L_{7}^{\circ}=0$, $L_{6}^{\circ}=
-\Lambda$.

\section{All $H$-potentials of Lanczos potentials of the Weyl spinor 
that are aligned to $o^{A'}$}

We will say that a completely symmetric spinor $H_{ABA'B'}$ is aligned 
to $o^{A'}$ if it has the algebraic structure $H_{ABA'B'}=Q_{AB}o_{A'}
o_{B'}$. In this section we will find all such spinors $H_{ABA'B'}$ that
are solutions of the equation
\begin{equation}
	L_{ABCA'}=\nabla_{(A}{}^{B'}H_{BC)A'B'}
	\label{LHeqn}
\end{equation}
where $L_{ABCA'}$ is a Lanczos potential of the Weyl spinor, i.e.,
$$
 \Psi_{ABCD}=2\nabla_{(A}{}^{A'}L_{BCD)A'},
$$
in spacetimes of class ${\cal G}$ with spinor dyad in standard form.

First we note that if $H_{ABA'B'}$ is aligned to $o^{A'}$ and satisfies
(\ref{LHeqn}) then
\begin{eqnarray}
 L_{ABCA'}o^{A'} & = & o^{A'}\nabla_{(A}{}^{B'}H_{BC)A'B'}=-Q_{(BC}o^{A'}
 o^{B'}\nabla_{A)B'}o_{A'} \nonumber \\ & = & \bar{\kappa}Q_{(BC}\iota_{A)}
 -\bar{\sigma}Q_{(BC}o_{A)}=0 \nonumber
\end{eqnarray}
so that $L_{ABCA'}$ has the algebraic structure $L_{ABCA'}=M_{ABC}
o_{A'}$ for some symmetric spinor $M_{ABC}$ and is therefore itself 
aligned to $o^{A'}$. Hence, it suffices to solve equation (\ref{LHeqn})
for the Lanczos potentials found in the previous section. We remark 
that since the spacetimes we are considering are not necessarily 
Einstein, and since we are only considering $H$-potentials that are 
aligned to $o^{A'}$, their existence is {\em not} guaranteed by the 
results in \cite{AE2}. 

If $H_{ABA'B'}$ is aligned to $o^{A'}$ it follows that only the components
$H_{02'}$, $H_{12'}$ and $H_{22'}$ are non-zero and from the above 
calculation we see that four out of the eight GHP-equations are identically
satisfied. The remaining four become, using Held's operators
\begin{eqnarray}
 L_{4} & = & -\Ph H_{02'}+\bar{\rho}H_{02'} \nonumber \\Ê3L_{5} & = &
 -2\Ph H_{12'}-\rho\Edt' H_{02'}-2(\rho-\bar{\rho})H_{12'} \nonumber \\
 3L_{6} & = & -\Ph H_{22'}-2\rho\Edt'H_{12'}-(2\rho-\bar{\rho})H_{22'}
 \nonumber \\ L_{7} & = & -\rho\Edt' H_{22'} \label{GHPLH} 
\end{eqnarray}
The first three of these equations can now be $\rho$-integrated in the same
way as in the previous section and after some calculations we obtain
\begin{eqnarray}
 H_{02'} & = & \frac{\bar{\rho}}{\rho}L_{4}^{\circ}+\bar{\rho}
 H_{02'}^{\circ} \nonumber \\ H_{12'} & = & \frac{3}{2}
 \frac{\bar{\rho}}{\rho^2}L_{5}^{\circ}+\frac{\bar{\rho}}{\rho}
 H_{12'}^{\circ}-\frac{1}{2}\bar{\rho}\bigl(\Edt' H_{02'}^{\circ}-
 L_{4}^{\circ}\Edt' \Omega^{\circ}\bigr) \nonumber \\ H_{22'} & = &
 3\frac{\bar{\rho}}{\rho^3}L_{6}^{\circ}+\frac{\bar{\rho}}{\rho^2}
 H_{22'}^{\circ}-\frac{1}{2}\frac{\bar{\rho}}{\rho}\bigl(4\Edt'
 H_{12'}^{\circ}+\Edt'^2 L_{4}^{\circ}-9L_{5}^{\circ}\Edt'\Omega^{\circ}
 \bigr)+\frac{1}{4}\rho \Psi_{2}^{\circ} \nonumber \\ & & +\frac{1}{2}
 \bar{\rho}\bigl(\Edt'^2 H_{02'}^{\circ}+2H_{12'}^{\circ}\Edt'\Omega^{\circ}
 -L_{4}^{\circ}\Edt'^2 \Omega^{\circ}-\Edt'\Omega^{\circ}\Edt' L_{4}^{\circ}
 +\frac{1}{2}\Psi_{2}^{\circ}\bigr) \nonumber \\Ê& & +\frac{1}{2}\rho
 \bar{\rho}\Phi_{11}^{\circ}
 \label{}
\end{eqnarray}
These $H$-scalars now give an $H$-potential of a Lanczos potential of 
the Weyl spinor if and only if the last equation of (\ref{GHPLH}) is 
satisfied. By substituting the above expressions for $L_{7}$ and 
$H_{22'}$ into this equation we find that it is satisfied if and only
if
\begin{eqnarray}
0  & = & L_{7}^{\circ}+\rho^2 \bigl(\Edt' H_{22'}^{\circ}+\frac{3}{2}
\Edt' L_{5}^{\circ}-6L_{6}^{\circ}\Edt' \Omega^{\circ}\bigr) \nonumber
\\ & & -2\rho^3 \bigl(\Edt'^2 H_{12'}^{\circ}+H_{22'}^{\circ}\Edt'
\Omega^{\circ}+\frac{1}{3}\Edt'^3 L_{4}^{\circ}-2L_{5}^{\circ}\Edt'^2
\Omega^{\circ}-\frac{3}{2}\Edt' \Omega^{\circ}\Edt' L_{5}^{\circ}+
\frac{1}{3}\Psi_{3}^{\circ}\bigr) \nonumber \\ & & +\frac{1}{2}\rho^4 \bigl(
\Edt'^3 H_{02'}^{\circ}+2H_{12'}^{\circ}\Edt'^2 \Omega^{\circ}+6\Edt'
\Omega^{\circ}\Edt' H_{12'}^{\circ}-L_{4}^{\circ}\Edt'^3 \Omega^{\circ}-2
\Edt'^2 \Omega^{\circ}\Edt' L_{4}^{\circ} \nonumber \\ & & -9L_{5}^{\circ}
(\Edt' \Omega^{\circ})^2+\frac{1}{2}\Edt' \Psi_{2}^{\circ}-\Omega^{\circ}
\Psi_{3}^{\circ}-\Phi_{21}^{\circ}\bigr)
\end{eqnarray}
By repeatedly taking $\Ph$ of the above equation and dividing by 
$\rho^2$ we obtain the following necessary and sufficient conditions 
for $H_{ABA'B'}$, aligned to $o^{A'}$, to be an $H$-potential of a 
Lanczos potential of the Weyl spinor.
\begin{eqnarray}
 0 & = & L_{7}^{\circ} \nonumber \\Ê0 & = & \Edt' H_{22'}^{\circ}+\frac{3}{2}
 \Edt' L_{5}^{\circ}-6L_{6}^{\circ}\Edt' \Omega^{\circ} \nonumber \\ 
 0 & = & \Edt'^2 H_{12'}^{\circ}+H_{22'}^{\circ}\Edt' \Omega^{\circ}+
 \frac{1}{3}\Edt'^3 L_{4}^{\circ}-2L_{5}^{\circ}\Edt'^2 \Omega^{\circ}-
 \frac{3}{2}\Edt' \Omega^{\circ}\Edt' L_{5}^{\circ}+\frac{1}{3}\Psi_{3}^{\circ}
 \nonumber \\ 0 & = & \Edt'^3 H_{02'}^{\circ}+2H_{12'}^{\circ}\Edt'^2
 \Omega^{\circ}+6\Edt' \Omega^{\circ}\Edt' H_{12'}^{\circ}-L_{4}^{\circ}\Edt'^3
 \Omega^{\circ}-2\Edt'^2 \Omega^{\circ}\Edt' L_{4}^{\circ} \nonumber \\ & &
 -9L_{5}^{\circ}(\Edt' \Omega^{\circ})^2+\frac{1}{2}\Edt' \Psi_{2}^{\circ}-
 \Omega^{\circ}\Psi_{3}^{\circ}-\Phi_{21}^{\circ}
 \label{HpotConditions}
\end{eqnarray}
Now, because $\Bigl[\Ph,\Edt'\Bigr]=0$ the second of these equations 
must have a local solution, $H_{22'}^{\circ}$. By substituting this 
solution into the third equation, a local solution, $H_{12'}^{\circ}$,
of this equation must exist, and similarly the fourth equation must
have a local solution $H_{02'}^{\circ}$. Thus, a Lanczos potential of 
the Weyl spinor has an $H$-potential that is aligned to $o^{A'}$ 
if and only if $L_{7}^{\circ}=0$.

Summing up, we have proved the following result:
\begin{theorem} \label{Hpotentials}
 For any spacetime of class ${\cal G}$ with spinor dyad in standard 
 form, all $H$-potentials that are aligned to $o^{A'}$, of Lanczos
 potentials of the Weyl spinor, are given by
 \begin{eqnarray}
  H_{02'} & = & \frac{\bar{\rho}}{\rho}L_{4}^{\circ}+\bar{\rho}H_{02'}^{\circ}
  \nonumber \\ H_{12'} & = & \frac{3}{2}\frac{\bar{\rho}}{\rho^2}L_{5}^{\circ}
  +\frac{\bar{\rho}}{\rho}H_{12'}^{\circ}-\frac{1}{2}\bar{\rho}\bigl(\Edt'
  H_{02'}^{\circ}-L_{4}^{\circ}\Edt' \Omega^{\circ}\bigr) \nonumber \\
  H_{22'} & = & 3\frac{\bar{\rho}}{\rho^3}L_{6}^{\circ}+\frac{\bar{\rho}}
  {\rho^2}H_{22'}^{\circ}-\frac{1}{2}\frac{\bar{\rho}}{\rho}\bigl(4\Edt'
  H_{12'}^{\circ}+\Edt'^2 L_{4}^{\circ}-9L_{5}^{\circ}\Edt' \Omega^{\circ}
  \bigr)+\frac{1}{4}\rho \Psi_{2}^{\circ} \nonumber \\ & & +\frac{1}{2}
  \bar{\rho}\bigl(\Edt'^2 H_{02'}^{\circ}+2H_{12'}^{\circ}\Edt' \Omega^{\circ}
  -L_{4}^{\circ}\Edt'^2 \Omega^{\circ}-\Edt' \Omega^{\circ}\Edt' L_{4}^{\circ}
  +\frac{1}{2}\Psi_{2}^{\circ}\bigr) \nonumber \\Ê& & +\frac{1}{2}\rho\bar{\rho}
  \Phi_{11}^{\circ}. \label{allHpot}
 \end{eqnarray}
 The functions of integration $L_{4}^{\circ}, L_{5}^{\circ}, L_{6}^{\circ},
 H_{02'}^{\circ}, H_{12'}^{\circ}$ and $H_{22'}^{\circ}$ are subject to
 the conditions
 \begin{eqnarray}
   0 & = & \Edt'^2 L_{6}^{\circ} \nonumber \\ 0 & = & \Edt'^3 L_{5}^{\circ}
   +2L_{6}^{\circ}\Edt'^2 \Omega^{\circ}+6\Edt' \Omega^{\circ}\Edt'
   L_{6}^{\circ}+\frac{1}{3}\Psi_{4}^{\circ} \nonumber \\ 0 & = & \Edt'^4
   L_{4}^{\circ}+3L_{5}^{\circ}\Edt'^3 \Omega^{\circ}+12\Edt'^2
   \Omega^{\circ}\Edt' L_{5}^{\circ}+18\Edt' \Omega^{\circ}\Edt'^2
   L_{5}^{\circ}+18L_{6}^{\circ}(\Edt' \Omega^{\circ})^2 \nonumber \\ & &
   +\Edt' \Psi_{3}^{\circ}-3\Omega^{\circ}\Psi_{4}^{\circ} \nonumber 
   \\ 0 & = & \Edt' H_{22'}^{\circ}+\frac{3}{2}\Edt'^2 L_{5}^{\circ}-6
   L_{6}^{\circ}\Edt' \Omega^{\circ} \nonumber \\ 0 & = & \Edt'^2
   H_{12'}^{\circ}+H_{22'}^{\circ}\Edt' \Omega^{\circ}+\frac{1}{3}\Edt'^3
   L_{4}^{\circ}-2L_{5}^{\circ}\Edt'^2 \Omega^{\circ}-\frac{3}{2}\Edt'
   \Omega^{\circ}\Edt' L_{5}^{\circ}+\frac{1}{3}\Psi_{3}^{\circ} \nonumber
   \\ 0 & = & \Edt'^3 H_{02'}^{\circ}+2H_{12'}^{\circ}\Edt'^2 \Omega^{\circ}
   +6\Edt' \Omega^{\circ}\Edt' H_{12'}^{\circ}-L_{4}^{\circ}\Edt'^3
   \Omega^{\circ}-2\Edt'^2 \Omega^{\circ}\Edt' L_{4}^{\circ} \nonumber \\ & &
   -9L_{5}^{\circ}(\Edt' \Omega^{\circ})^2+\frac{1}{2}\Edt' \Psi_{2}^{\circ}-
   \Omega^{\circ}\Psi_{3}^{\circ}-\Phi_{21}^{\circ} 
   \label{LHConditions}
 \end{eqnarray}
 and in particular, there always exists a local $H$-potential that is 
 aligned to $o^{A'}$.
\end{theorem}
We also note that the Lanczos scalars of the Lanczos potentials obtained
in this theorem, are given by (\ref{Lscalars}) with 
$L_{7}^{\circ}=0$ and for future reference, we also note that a simple
particular solution of the first equation is $L_{6}^{\circ}=-\Lambda$. 

\section{Lanczos potentials and curvature-free connections}

\subsection{Riemann-Cartan equations}

It is well-known \cite{PR1} that given any spinor $\Gamma_{ABCA'}=
\Gamma_{(AB)CA'}$ we can define a metric connection 
$\hat{\nabla}_{AA'}$ by the equation
\begin{equation}
	\hat{\nabla}_{AA'}\xi^{B}=\nabla_{AA'}\xi^{B}+2\Gamma_{C}{}^{B}
	{}_{AA'}\xi^{C}
	\label{}
\end{equation}
and providing $\Gamma_{ABCA'}\neq 0$ the connection 
$\hat{\nabla}_{AA'}$ will have non-zero torsion. The curvature of 
such a connection can be described by its curvature spinors
$\hat{\Psi}_{ABCD}=\hat{\Psi}_{(ABCD)}$, $\hat{\Phi}_{ABA'B'}=
\hat{\Phi}_{(AB)(A'B')}$, $\hat{\Sigma}_{AB}=\hat{\Sigma}_{(AB)}$
and $\hat{\Lambda}$, through the formula \cite{Andersson}, \cite{AE1}
\begin{eqnarray}
 \hat{R}_{abcd} & = & \eps_{A'B'}\eps_{C'D'}\bigl[\hat{\Psi}_{ABCD}+
 2(\eps_{B(C}\hat{\Sigma}_{D)A}+\eps_{A(C}\hat{\Sigma}_{D)B}) 
 \nonumber \\Ê& & +\hat{\Lambda}(\eps_{AD}\eps_{BC}+\eps_{AC}\eps_{BD})
 \bigr]+\hat{\Phi}_{ABC'D'}\eps_{A'B'}\eps_{CD} \nonumber \\Ê& & +c.c
\end{eqnarray}
where $c.c$ stands for the complex conjugate of the entire expression.

Note that if the torsion is non-zero then $\hat{\Phi}_{ABA'B'}$ and
$\hat{\Lambda}$ are in general complex quantities and $\hat{\Sigma}_{AB}$
is in general non-zero.

The curvature spinors of $\hat{\nabla}_{AA'}$ are related to the curvature
spinors of $\nabla_{AA'}$, \cite{Andersson}, \cite{AE1}
\begin{eqnarray}
 \hat{\Psi}_{ABCD} & = & \Psi_{ABCD}-2\nabla_{(A}{}^{E'}\Gamma_{BCD)E'}
 -4\Gamma_{E(AB}{}^{E'}\Gamma^{E}{}_{CD)E'} \nonumber \\ \hat{\Lambda}
 & = & \Lambda-\frac{1}{3}\nabla_{E}{}^{E'}\Gamma^{EF}{}_{FE'}-
 \frac{1}{3}\Gamma_{EFGE'}\Gamma^{EGFE'}+\frac{1}{3}\Gamma^{F}{}_{EFE'}
 \Gamma^{EG}{}_{G}{}^{E'} \nonumber \\ \hat{\Sigma}_{AB} & = &
 \frac{1}{4}\nabla^{EE'}\Gamma_{E(AB)E'}-\frac{1}{4}\nabla_{(A}{}^{E'}
 \Gamma^{E}{}_{B)EE'}-\frac{1}{2}\Gamma_{E(A|F|}{}^{E'}\Gamma^{EF}
 {}_{B)E'} \nonumber \\Ê& & -\frac{1}{2}\Gamma_{E(AB)}{}^{E'}
 \Gamma^{EF}{}_{FE'} \nonumber \\Ê\hat{\Phi}_{ABA'B'} & = & \Phi_{ABA'B'}
 -2\nabla_{(A}{}^{E'}\bar{\Gamma}_{|A'B'E'|B)}+4\bar{\Gamma}_{A'E'F'(A}
 \bar{\Gamma}_{|B'|}{}^{E'F'}{}_{B)}
\end{eqnarray}
Now, $\Gamma_{ABCA'}$ can be decomposed into a symmetric (3,1)-spinor
$L_{ABCA'}$ and a complex covector $\ch_{AA'}$ according to
\begin{equation}
	\Gamma_{ABCA'}=L_{ABCA'}+\eps_{AC}\ch_{BA'}+\eps_{BC}\ch_{AA'}
	\label{}
\end{equation}
where $L_{ABCA'}=\Gamma_{(ABC)A'}$ and $\ch_{AA'}=\frac{1}{3}
\Gamma_{AB}{}^{B}{}_{A'}$. It can then be shown \cite{Andersson} that
the curvature spinors of $\hat{\nabla}_{AA'}$ can be expressed as
\begin{eqnarray}
 \hat{\Psi}_{ABCD} & = & \Psi_{ABCD}-2\nabla_{(A}{}^{E'}L_{BCD)E'}-8
 \ch_{(A}{}^{E'}L_{BCD)E'}+4L_{(AB}{}^{EE'}L_{CD)EE'} \nonumber \\
 \hat{\Lambda} & = & \Lambda-\nabla^{EE'}\ch_{EE'}-\frac{1}{3}L_{EFGE'}
 L^{EFGE'}+4\ch_{EE'}\ch^{EE'} \nonumber \\ \hat{\Sigma}_{AB} & = &
 \frac{1}{4}\nabla^{EE'}L_{ABEE'}+\nabla_{(A}{}^{E'}\ch_{B)E'}-3
 L_{AB}{}^{EE'}\ch_{EE'} \nonumber \\ \hat{\Phi}_{ABA'B'} & = & \Phi_{ABA'B'}
 -2\nabla_{(A}{}^{E'}\bar{L}_{|A'B'E'|B)}+2\nabla_{(A|A'}
 \bar{\ch}_{B'|B)}+2\nabla_{(A|B'}\bar{\ch}_{A'|B)} \nonumber \\ & &
 +4\bar{L}_{A'E'F'(A}\bar{L}_{|B'|}{}^{E'F'}{}_{B)}+8\bar{L}_{A'B'E'(A}
 \bar{\ch}^{E'}{}_{B)} \nonumber \\ & & +16\bar{\ch}_{A'(A}
 \bar{\ch}_{|B'|B)} \label{DecompRhat}
\end{eqnarray}
We note that the corresponding equation in both \cite{Andersson} and
\cite{AE1} unfortunately contains a misprint in the coefficient of the
last term. These equations will be used to find connections on the
spacetimes studied in the previous sections, that are curvature-free and
for which $L_{ABCA'}$ is a Lanczos potential of the Weyl
spinor that is aligned to $o^{A'}$.

\subsection{Kerr-Schild spacetimes, Lanczos potentials, curvature-free
connections and quasi-local momentum} 

In \cite{BL1} Bergqvist and Ludvigsen study the Kerr spacetime. It is 
known to be a special case of a Kerr-Schild spacetime, i.e., its metric 
can be written
\begin{equation}
	g_{ab}=\eta_{ab}+2fl_{a}l_{b}
	\label{}
\end{equation}
where $\eta_{ab}$ is a flat metric, $l^{a}=o^{A}o^{A'}$ is a null vector
that, in the Kerr case, is geodesic and shear-free and $f$ is a real 
function that can be written
\begin{equation}
	f=\frac{\rho+\bar{\rho}}{4\rho^3}\Psi_{2}
	\label{}
\end{equation}
in the Kerr case. If we put
\begin{equation}
	H_{ABA'B'}=fo_{A}o_{B}o_{A'}o_{B'}=fl_{a}l_{b}
	\label{}
\end{equation}
it was shown in \cite{BL1} that the spinor
$$
 \Gamma_{ABCA'}=\nabla_{(A}{}^{B'}H_{B)CA'B'}
$$
defines a metric connection with non-zero torsion, but vanishing 
curvature, i.e., $\hat{R}_{abcd}=0$. In \cite{Bergqvist} it was 
subsequently shown that the spinor
$$
 L_{ABCA'}=\Gamma_{(ABC)A'}=\nabla_{(A}{}^{B'}H_{BC)A'B'}
$$
is a Lanczos potential of the Weyl spinor that is aligned to $o^{A'}$.

These results were generalized in \cite{Harnett} and \cite{AE1}. The 
final result is that in any Kerr-Schild spacetime where $l^{a}=o^{A}
o^{A'}$ is geodesic  and shear-free, the above construction yields a
metric, asymmetric, curvature-free connection $\hat{\nabla}_{AA'}$ with
the property that $L_{ABCA'}=\Gamma_{(ABC)A'}$ is a Lanczos potential
of the Weyl spinor that is aligned to $o^{A'}$.

In \cite{BL}, \cite{BL1} Bergqvist and Ludvigsen used the curvature-free
connection $\nabla_{AA'}$ described previously, to define quasi-local
momentum in the Kerr spacetime. In this section we will review this
construction.

That $\hat{\nabla}_{AA'}$ is curvature-free means that it is 
integrable, i.e., parallel propagation is path independent. From this 
fact we can easily prove that the spinor fields that satisfy the equation
\begin{equation}
	\hat{\nabla}_{AA'}\xi_{B}=0
	\label{}
\end{equation}
form a 2-dimensional vector space over the complex numbers. We will 
call this vector space of spinor fields ${\cal S}$ (with indices 
according to the abstract index notation \cite{PR1} when appropriate). For
a spinor field $\xi_{A}\in {\cal S}_{A}$ we define the spinor
\begin{equation}
	\varphi_{AB}=\xi_{(A}\nabla_{B)}{}^{C'}\bar{\xi}_{C'}-\bar{\xi}_{C'}
	\nabla_{(A}{}^{C'}\xi_{B)}
	\label{}
\end{equation}
and the (antisymmetric) 2-form
\begin{equation}
    F_{ab}=i\bigl(\eps_{AB}\bar{\varphi}_{A'B'}-\eps_{A'B'}\varphi_{AB}
    \bigr). \label{}
\end{equation}
Bergqvist and Ludvigsen prove that $F_{ab}$ is actually a closed 2-form,
i.e., $\nabla_{[a}F_{bc]}=0$. Given a spacelike 2-surface $\Sigma$ they
then define the quasi-local momentum $P_{AA'}(\Sigma)$ as a 1-form on
the hermitian part of ${\cal S}^{A}\otimes\bar{{\cal S}}^{A'}$, by the
equation
\begin{equation}
    P_{AA'}(\Sigma)\xi^{A}\bar{\xi}^{A'}=\frac{1}{8\pi}\int_{\Sigma}
    {\bf F}.
    \label{}
\end{equation}
This defines the action of $P_{AA'}(\Sigma)$ on null vector fields in
the hermitian part of ${\cal S}^{A}\otimes \bar{{\cal S}}^{A'}$ and by
linearity its action is then defined on all of the hermitian part of
${\cal S}^{A}\otimes\bar{{\cal S}}^{A'}$. We note that this definition is
genuinely quasi-local as we have made no reference to the asymptotic
properties of the Kerr spacetime. $P_{AA'}(\Sigma)$ can also be shown to,
in a certain sense, agree with the Bondi momentum when $\Sigma$ is a cross 
section of future null infinity.

\subsection{Connections and Lanczos potentials in class ${\cal G}$ 
spacetimes that are aligned to $o^{A'}$}

\subsubsection{Connections for which $\hat{\Psi}_{ABCD}=0$,
$\hat{\Sigma}_{AB}=0$}

We will now give a similar construction using the Lanczos 
potentials and $H$-potentials that were found in the previous 
sections, as in the Kerr-Schild case. Thus, suppose once again
that we have an arbitrary class ${\cal G}$ spacetime with spinor dyad 
in standard form.

If we choose $H_{ABA'B'}$ to be aligned to $o^{A'}$ then, as is already
shown, $L_{ABCA'}$ will automatically be aligned to $o^{A'}$. In a similar
way, it is easy to show that $\ch_{AA'}=\lambda_{A}o_{A'}$ for some spinor
$\lambda_{A}$. It automatically follows that all the product terms in the
first three equations of (\ref{DecompRhat}) vanish. Moreover, if we choose
$H_{ABA'B'}$ as in Theorem \ref{Hpotentials}, so that $L_{ABCA'}$ is a
Lanczos potential of the Weyl spinor, it is easily seen that
$\hat{\Psi}_{ABCD}=0$. Hence, we immediately get the result
\begin{proposition}
 Let $H_{ABA'B'}$ be as in Theorem \ref{Hpotentials}. Then the spinor
 $$
  \Gamma_{ABCA'}=\nabla_{(A}{}^{B'}H_{B)CA'B'}
 $$
 defines a metric connection $\hat{\nabla}_{AA'}$ through the equation
 $$
  \hat{\nabla}_{AA'}\xi^{B}=\nabla_{AA'}\xi^{B}+2\Gamma_{C}{}^{B}
  {}_{AA'}\xi^{C}
 $$
 that is $\hat{\Psi}$-flat, i.e., $\hat{\Psi}_{ABCD}=0$. 
\end{proposition}
We will next choose a particular class of $H$-potentials that will 
ensure that the curvature spinor $\hat{\Sigma}_{AB}$ vanishes. We will do
this in two steps. First we will $\rho$-integrate the GHP-version of the
corresponding equations from (\ref{DecompRhat}) to get $\ch_{AA'}$. Then
we note that from the definition of $\ch_{AA'}$ we have
\begin{equation}
	\ch_{AA'}=\frac{1}{3}\Gamma_{AB}{}^{B}{}_{A'}=-\frac{1}{6}\nabla^{BB'}
	H_{ABA'B'},
	\label{}
\end{equation}
so we then substitute our expressions for the various quantities into 
the GHP-version of this equation to get the possible choices for
$H_{ABA'B'}$.

Hence, first we wish to solve the equations
\begin{equation}
 0=\hat{\Sigma}_{AB}=\frac{1}{4}\nabla^{EE'}L_{ABEE'}+\nabla_{(A}{}^{E'}
 \ch_{B)E'} \label{zeroSigma}
\end{equation}
We note that since by assumption $\ch_{AA'}=\lambda_{A}o_{A'}$ it 
has only two non-vanishing components, namely
\begin{eqnarray*}
 \ch_{01'} & = & \ch_{AA'}o^{A}\iota^{A'} \\ \ch_{11'} & = & 
 \ch_{AA'}\iota^{A}\iota^{A'}
\end{eqnarray*}
Then the GHP-version of (\ref{zeroSigma}) becomes
\begin{eqnarray}
 0 & = & -\Ph \ch_{01'}+\bar{\rho}\ch_{01'}+\frac{1}{4}\bigl(\Ph L_{5}
 -\rho\Edt' L_{4}-(3\rho+\bar{\rho})L_{5}\bigr) \\ 0 & = & -\Ph \ch_{11'}
 -\rho\Edt' \ch_{01'}-(\rho-\bar{\rho})\ch_{11'}+\frac{1}{2}\bigl(
 \Ph L_{6}-\rho\Edt' L_{5}-(2\rho+\bar{\rho})L_{6}\bigr) \\ 0 & = & -\rho\Edt'
 \ch_{11'}+\frac{1}{4}\bigl(\Ph L_{7}-\rho\Edt' L_{6}-(\rho+\bar{\rho})L_{7}
 \bigr) \label{GHPchL}
\end{eqnarray}
By using the Weyl-Lanczos equations we can eliminate $\Edt' L_{i}, i=4,5,6$
from the above equations and by substituting the expressions from Section 3
for the Lanczos scalars it is possible to $\rho$-integrate the first two of
these equations,
\begin{eqnarray}
 \ch_{01'} & = & \frac{\bar{\rho}}{\rho}L_{5}^{\circ}+\bar{\rho}
 \ch_{01'}^{\circ} \nonumber \\ \ch_{11'} & = & 2\frac{\bar{\rho}}
 {\rho^2}L_{6}^{\circ}+\frac{\bar{\rho}}{\rho}\ch_{11'}^{\circ}
 -\bar{\rho}\bigl(\Edt' \ch_{01'}^{\circ}-L_{5}^{\circ}\Edt' \Omega^{\circ}
 \bigr)+\frac{1}{12}\rho\bar{\rho}\Psi_{2}^{\circ}. \label{chiscalars}
\end{eqnarray}
We now need to substitute this into the third equation, but before we 
do that we will temporarily drop the assumption that $\ch_{AA'}=-\frac{1}{6}
\nabla^{BB'}H_{ABA'B'}$ and instead just assume that $\ch_{AA'}=\lambda_{A}
o_{A'}$ so that we allow for a non-zero $L_{7}^{\circ}$. Then the third
equation becomes
\begin{eqnarray}
 0 & = & L_{7}^{\circ}+\rho^2\bigl(\Edt' \ch_{11'}^{\circ}+\frac{1}{2}
 \Edt'^2 L_{5}^{\circ}-3L_{6}^{\circ}\Edt' \Omega^{\circ}\bigr) 
 \nonumber \\Ê& & -\rho^3 \bigl(\Edt'^2 \ch_{01'}^{\circ}+\ch_{11'}^{\circ}
 \Edt' \Omega^{\circ}-L_{5}^{\circ}\Edt'^2 \Omega^{\circ}-\Edt' \Omega^{\circ}
 \Edt' L_{5}^{\circ}+\frac{1}{6}\Psi_{3}^{\circ}\bigr)
\end{eqnarray}
By identifying coefficients in the same way as in the previous 
sections we obtain the conditions
\begin{eqnarray}
 0 & = & L_{7}^{\circ} \nonumber \\ 0 & = & \Edt' \ch_{11'}^{\circ}+
 \frac{1}{2}\Edt'^2 L_{5}^{\circ}-3L_{6}^{\circ}\Edt' \Omega^{\circ}
 \nonumber \\Ê0 & = & \Edt'^2 \ch_{01'}^{\circ}+\ch_{11'}^{\circ}
 \Edt' \Omega^{\circ}-L_{5}^{\circ}\Edt'^2 \Omega^{\circ}-\Edt'
 \Omega^{\circ}\Edt' L_{5}^{\circ}+\frac{1}{6}\Psi_{3}^{\circ}
 \label{chiConditions}
\end{eqnarray}
By the commutator $\Bigl[\Ph, \Edt'\Bigr]=0$ it follows that we can 
solve the second of these equations for $\ch_{11'}^{\circ}$, 
substitute the result into the third equation and solve it for
$\ch_{01'}^{\circ}$. Hence, it follows that we can choose $\ch_{AA'}$ 
so that $\hat{\Sigma}_{AB}=0$ if and only if $L_{7}^{\circ}=0$. 
Recall from the previous section that our Lanczos potential $L_{ABCA'}$
possessed an $H$-potential if and only if $L_{7}^{\circ}=0$ so the 
Lanczos potentials that allow us to obtain a connection of the above 
type, with $\hat{\Sigma}_{AB}=0$ are precisely the Lanczos potentials
that possess an $H$-potential that is aligned to $o^{A'}$. However, it
remains to be seen whether the $H$-potential can be chosen so that
$\ch_{AA'}=-\frac{1}{6}\nabla^{BB'}H_{ABA'B'}$, i.e., so that
$$
  \Gamma_{ABCA'}=\nabla_{(A}{}^{B'}H_{B)CA'B'}.
$$
This will be the topic of our next investigation.

The GHP-version of the equation $\ch_{AA'}=-\frac{1}{6}\nabla^{BB'}
H_{ABA'B'}$, as two of the four equations are identically satisfied, is
\begin{eqnarray}
 6\ch_{01'} & = & -\Ph H_{12'}+\rho\Edt' H_{02'}+(2\rho+\bar{\rho})H_{12'}
 \\ 6\ch_{11'} & = & -\Ph H_{22'}+\rho\Edt' H_{12'}+(\rho+\bar{\rho})H_{22'}
\end{eqnarray}
We can use the equations (\ref{GHPLH}) to eliminate the quantities 
$\Edt' H_{02'}$ and $\Edt' H_{12'}$ from these equations and we can
substitute the expressions for the Lanczos-, $H$- and $\ch$-scalars
obtained previously, into these equations. Then they become, after some
simplification
\begin{eqnarray*}
 0 & = & 2\ch_{01'}^{\circ}-H_{12'}^{\circ}-\frac{1}{3}\Edt' L_{4}^{\circ}
 \\ 0 & = & 2\ch_{11'}^{\circ}-H_{22'}^{\circ}-\frac{1}{2}\Edt' L_{5}^{\circ}
 -\rho\bigl(2\Edt' \ch_{01'}^{\circ}-\Edt' H_{12'}^{\circ}-\frac{1}{3}\Edt'^2
 L_{4}^{\circ}\bigr)
\end{eqnarray*}
We see that if the first of these conditions is satisfied, then the 
expression within parenthesis in the second is identically zero. 
Hence, the conditions simplify to
\begin{eqnarray}
 \ch_{01'}^{\circ} & = & \frac{1}{2}H_{12'}^{\circ}+\frac{1}{6}\Edt'
 L_{4}^{\circ} \nonumber \\ \ch_{11'}^{\circ} & = & \frac{1}{2}
 H_{22'}^{\circ}+\frac{1}{4}\Edt' L_{5}^{\circ} \label{HchiConditions}
\end{eqnarray}
We have chosen the $H$-scalars to satisfy (\ref{HpotConditions}).
Thus, we need to check that the $\ch$-scalars defined by 
(\ref{HchiConditions}) satisfy (\ref{chiConditions}). We obtain, 
according to (\ref{HpotConditions})
\begin{eqnarray*}
 0 & = & \frac{1}{2}\bigl(\Edt' H_{22'}^{\circ}+\frac{3}{2}\Edt'^2
 L_{5}^{\circ}-6L_{6}^{\circ}\Edt'\Omega^{\circ}\bigr) \\ & = & \Edt'\bigl(
 \frac{1}{2}H_{22'}^{\circ}+\frac{1}{4}\Edt' L_{5}^{\circ}\bigr)+\frac{1}{2}
 \Edt'^2 L_{5}^{\circ}-3L_{6}^{\circ}\Edt' \Omega^{\circ} \\ & = & \Edt'
 \ch_{11'}^{\circ}+\frac{1}{2}\Edt'^2 L_{5}^{\circ}-3L_{6}^{\circ}\Edt'
 \Omega^{\circ}
\end{eqnarray*}
which is precisely (\ref{chiConditions}). For $\ch_{01'}^{\circ}$ we 
obtain
\begin{eqnarray*}
 0 & = & \frac{1}{2}\bigl(\Edt'^2 H_{12'}^{\circ}+H_{22'}^{\circ}\Edt'
 \Omega^{\circ}-2L_{5}^{\circ}\Edt'^2 \Omega^{\circ}-\frac{3}{2}\Edt'
 \Omega^{\circ}\Edt' L_{5}^{\circ}+\frac{1}{3}\Edt'^3 L_{4}^{\circ}+
 \frac{1}{3}\Psi_{3}^{\circ}\bigr)\\ & = & \Edt'^2\bigl(\frac{1}{2}
 H_{12'}^{\circ}+\frac{1}{6}\Edt' L_{4}^{\circ}\bigr)+\bigl(\frac{1}{2}
 H_{22'}^{\circ}+\frac{1}{4}\Edt' L_{5}^{\circ}\bigr)\Edt' \Omega^{\circ}-
 L_{5}^{\circ}\Edt'^2 \Omega^{\circ} \\ & & -\Edt'\Omega^{\circ}\Edt'
 L_{5}^{\circ}+\frac{1}{6}\Psi_{3}^{\circ} \\Ê& = & \Edt'^2 \ch_{01'}^{\circ}
 +\ch_{11'}^{\circ}\Edt' \Omega^{\circ}-L_{5}^{\circ}\Edt'^2 \Omega^{\circ}-
 \Edt' \Omega^{\circ}\Edt' L_{5}^{\circ}+\frac{1}{6}\Psi_{3}^{\circ}
\end{eqnarray*}
which is also condition (\ref{chiConditions}).

The following result can now easily be proved:
\begin{theorem} \label{Sigmaflat}
 In spacetimes of class ${\cal G}$, a spinor $\Gamma_{ABCA'}=
 \Gamma_{(AB)CA'}$ that is aligned to $o^{A'}$ (i.e., $\Gamma_{ABCA'}=
 N_{ABC}o_{A'}$ with $N_{ABC}=N_{(AB)C}$) whose symmetric part $L_{ABCA'}
 =\Gamma_{(ABC)A'}$ is a Lanczos potential of the Weyl spinor, defines a
 connection $\hat{\nabla}_{AA'}$ for which $\hat{\Psi}_{ABCD}=0$ and
 $\hat{\Sigma}_{AB}=0$ if and only if it can be written
 $$
  \Gamma_{ABCA'}=\nabla_{(A}{}^{B'}H_{B)CA'B'}
 $$ 
 for some spinor $H_{ABA'B'}=Q_{AB}o_{A'}o_{B'}$ with 
 $Q_{AB}=Q_{(AB)}$. With a dyad in standard form, the Lanczos- and
 $H$-scalars for these spinors are given by (\ref{Lscalars}),
 (\ref{allHpot}) and (\ref{LHConditions}) with $L_{7}^{\circ}=0$.
 The $\ch$-scalars are given by (\ref{chiscalars}) and
 (\ref{HchiConditions}).
\end{theorem}

\begin{proof}
 Suppose the $\ch$- and $H$-scalars are related as in (\ref{HchiConditions}),
 i.e., $\ch_{AA'}=-\frac{1}{6}\nabla^{BB'}H_{ABA'B'}$. Then the above
 calculations prove that the conditions  (\ref{chiConditions}) and
 (\ref{HpotConditions}) are equivalent. Since (\ref{chiConditions}) 
 is equivalent to the vanishing of $\hat{\Sigma}_{AB}$ and since 
 (\ref{HpotConditions}) is equivalent to $H_{ABA'B'}$ being an 
 $H$-potential of a Lanczos potential of the Weyl spinor, the theorem 
 follows.
\end{proof}

We also note that, in particular, it follows that such spinors 
$\Gamma_{ABCA'}$ and $H_{ABA'B'}$ exist in every spacetime of class 
${\cal G}$. We remark that this partial result was proved in
\cite{Andersson} using a particular construction of Lanczos potentials
by Torres del Castillo. \cite{TdC1}, \cite{TdC2}.

\subsubsection{Connections for which $\hat{\Lambda}=0$}

We will now check whether our choice of $H$-potential also allows us to 
put $\hat{\Lambda}=0$. According to (\ref{DecompRhat}) the condition 
for this is
$$
 0=\Lambda-\nabla^{EE'}\ch_{EE'}=\Lambda-\Ph \ch_{11'}+\rho\Edt'\ch_{01'}
 +(\rho+\bar{\rho})\ch_{11'}.
$$
By using the expression (\ref{chiscalars}) for the $\ch$-scalars, we
arrive at the condition
$$
 0=L_{6}^{\circ}+\Lambda+\rho\bigl(2\ch_{11'}^{\circ}+\Edt' L_{5}^{\circ}
 +\Omega^{\circ}\Lambda\bigr)
$$
By identifying coefficients in the usual way we obtain the result 
that $\hat{\Lambda}=0$ if and only if
\begin{eqnarray}
 0 & = & L_{6}^{\circ}+\Lambda \nonumber \\Ê0 & = & 2\ch_{11'}^{\circ}
 +\Edt' L_{5}^{\circ}+\Omega^{\circ}\Lambda \label{lambdaConditions}
\end{eqnarray}
As remarked above, the first of these conditions satisfies the 
$L_{6}^{\circ}$-equation of (\ref{LpotConditions}) identically, as we 
have already chosen $L_{7}^{\circ}=0$ in order to get 
$\hat{\Sigma}_{AB}=0$ and in order to obtain an $H$-potential of
$L_{ABCA'}$ and we assumed that $\Lambda$ is constant. We now check the
second condition by substituting it into (\ref{chiConditions})
$$
 0=2\Edt'\ch_{11'}^{\circ}+\Edt'^2 L_{5}^{\circ}-6L_{6}^{\circ}
 \Edt' \Omega^{\circ}=\Edt'\bigl(-\Edt' L_{5}^{\circ}-\Omega^{\circ}
 \Lambda\bigr)+\Edt'^2 L_{5}^{\circ}+6\Lambda\Edt' \Omega^{\circ}=
 5\Lambda\Edt'\Omega^{\circ}.
$$
Hence, it is satisfied if and only if at least one of the conditions
$\Lambda=0$ and $\Edt'\Omega^{\circ}=0$ is satisfied. The second of 
these conditions is easily seen to be equivalent to the perhaps more 
familiar looking GHP-condition $\Ed'\rho=0$ which is satisfied, e.g., 
if $\rho=\bar{\rho}$.

If we now define $H$-scalars according to (\ref{HchiConditions}) it 
is clear that the conditions (\ref{HpotConditions}), for $H_{ABA'B'}$ 
to be an $H$-potential of $L_{ABCA'}$, are also identically satisfied 
if and only if $\Lambda=0$ or $\Edt'\Omega^{\circ}=0$.

Substituting (\ref{lambdaConditions}) into the equations 
(\ref{allHpot}) and (\ref{LHConditions}) and using that $\Lambda\Edt'
\Omega^{\circ}=0$ proves the following result:
\begin{lemma}
 Given a spacetime of class ${\cal G}$, there exists a spinor $H_{ABA'B'}
 =Q_{AB}o_{A'}o_{B'}$, $Q_{AB}=Q_{(AB)}$ such that the spinor
 $$
  \Gamma_{ABCA'}=\nabla_{(A}{}^{B'}H_{B)CA'B'}
 $$
 defines a metric, asymmetric connection for which
 $$
  \hat{\Psi}_{ABCD}=0\,,\;\hat{\Sigma}_{AB}=0\,,\;\hat{\Lambda}=0
 $$
 if and only if $\Lambda=0$ or $\Edt'\Omega^{\circ}=0$ 
 $(\Leftrightarrow \Ed'\rho=0)$. All such spinors $H_{ABA'B'}$ are 
 given by
 \begin{eqnarray}
  H_{02'} & = & \frac{\bar{\rho}}{\rho}L_{4}^{\circ}+\bar{\rho}H_{02'}^{\circ}
  \nonumber \\ H_{12'} & = & \frac{3}{2}\frac{\bar{\rho}}{\rho^2}L_{5}^{\circ}
  +\frac{\bar{\rho}}{\rho}H_{12'}^{\circ}-\frac{1}{2}\bar{\rho}\bigl(\Edt'
  H_{02'}^{\circ}-L_{4}^{\circ}\Edt' \Omega^{\circ}\bigr) \nonumber \\
  H_{22'} & = & -3\frac{\bar{\rho}}{\rho^3}\Lambda-\frac{\bar{\rho}}
  {\rho^2}\bigl(\frac{3}{2}\Edt' L_{5}^{\circ}+\Omega^{\circ}\Lambda\bigr)
  -\frac{1}{2}\frac{\bar{\rho}}{\rho}\bigl(4\Edt' H_{12'}^{\circ}+\Edt'^2
  L_{4}^{\circ}-9L_{5}^{\circ}\Edt' \Omega^{\circ}\bigr) \nonumber 
  \\ & & +\frac{1}{4}\rho\Psi_{2}^{\circ}+\frac{1}{2}\bar{\rho}\bigl(\Edt'^2
  H_{02'}^{\circ}+2H_{12'}^{\circ}\Edt' \Omega^{\circ}-L_{4}^{\circ}\Edt'^2
  \Omega^{\circ}-\Edt' \Omega^{\circ}\Edt' L_{4}^{\circ}+\frac{1}{2}
  \Psi_{2}^{\circ}\bigr) \nonumber \\Ê& & +\frac{1}{2}\rho\bar{\rho}
  \Phi_{11}^{\circ} \label{allHpot2}
 \end{eqnarray}
 where $L_{4}^{\circ}, L_{5}^{\circ}, H_{02'}^{\circ}$ and $H_{12'}^{\circ}$
 are subject to the conditions
 \begin{eqnarray}
   0 & = & \Edt'^3 L_{5}^{\circ}+\frac{1}{3}\Psi_{4}^{\circ} \nonumber \\
   0 & = & \Edt'^4
   L_{4}^{\circ}+3L_{5}^{\circ}\Edt'^3 \Omega^{\circ}+12\Edt'^2
   \Omega^{\circ}\Edt' L_{5}^{\circ}+18\Edt' \Omega^{\circ}\Edt'^2
   L_{5}^{\circ}+\Edt' \Psi_{3}^{\circ}-3\Omega^{\circ}\Psi_{4}^{\circ}
   \nonumber \\ 0 & = & \Edt'^2 H_{12'}^{\circ}+\frac{1}{3}\Edt'^3
   L_{4}^{\circ}-2L_{5}^{\circ}\Edt'^2 \Omega^{\circ}-3\Edt'
   \Omega^{\circ}\Edt' L_{5}^{\circ}+\frac{1}{3}\Psi_{3}^{\circ} \nonumber
   \\ 0 & = & \Edt'^3 H_{02'}^{\circ}+2H_{12'}^{\circ}\Edt'^2 \Omega^{\circ}
   +6\Edt' \Omega^{\circ}\Edt' H_{12'}^{\circ}-L_{4}^{\circ}\Edt'^3
   \Omega^{\circ}-2\Edt'^2 \Omega^{\circ}\Edt' L_{4}^{\circ} \nonumber \\ & &
   -9L_{5}^{\circ}(\Edt' \Omega^{\circ})^2+\frac{1}{2}\Edt' \Psi_{2}^{\circ}-
   \Omega^{\circ}\Psi_{3}^{\circ}-\Phi_{21}^{\circ} 
   \label{LHConditions2}
 \end{eqnarray}
\end{lemma}

\subsubsection{The Ricci spinor of $\hat{\nabla}_{AA'}$}

In this section we will consider the Ricci spinor of $\hat{\nabla}_{AA'}$.
We will therefore assume that $L_{ABCA'}$ and $\ch_{AA'}$ are both aligned
to $o^{A'}$ and have been chosen to give $\hat{\Psi}_{ABCD}=0$,
$\hat{\Sigma}_{AB}=0$, $\hat{\Lambda}=0$ in a spacetime of class ${\cal G}$
with dyad in standard form. Thus, in particular we assume that $\Lambda\Edt'
\Omega^{\circ}=0$, $L_{7}^{\circ}=0$, $L_{6}^{\circ}=-\Lambda$. Put
\begin{eqnarray}
 M_{ABC} & = & L_{ABCA'}\iota^{A'}=L_{7}o_{A}o_{B}o_{C}-3L_{6}o_{(A}
 o_{B}\iota_{C)}+3L_{5}o_{(A}\iota_{B}\iota_{C)}-L_{4}\iota_{A}\iota_{B}
 \iota_{C} \nonumber \\Ê\lambda_{A} & = & \ch_{AA'}\iota^{A'}=
 \ch_{11'}o_{A}-\ch_{01'}\iota_{A}
\end{eqnarray}
Then the complex conjugate of the fourth equation of (\ref{DecompRhat})
becomes
\begin{eqnarray}
	\bar{\hat{\Phi}}_{ABA'B'} & = & \Phi_{ABA'B'}-2M_{ABE}\nabla_{(A'}
	{}^{E}o_{B')}-2o_{(A'}\nabla_{B')}{}^{E}M_{ABE} \nonumber \\ & &
	+4\lambda_{(A}\nabla_{B)(A'}o_{B')}+4o_{(A'}\nabla_{B')(A}\lambda_{B)}
	+4M_{A}{}^{EF}M_{BEF}o_{A'}o_{B'} \nonumber \\Ê& & +8M_{ABE}\lambda^{E}
	o_{A'}o_{B'}+16\lambda_{A}\lambda_{B}o_{A'}o_{B'}
	\label{PhiMlambda}
\end{eqnarray}
Since $\hat{\Phi}_{ABA'B'}$ is in general non-hermitian it has 9 
complex components defined according to the usual convention \cite{PR1}.

Since 
$$
 o^{A'}o^{B'}\nabla_{AA'}o_{B'}=\bar{\sigma}o_{A}-\bar{\kappa}\iota_{A}=0,
$$
it follows from (\ref{PhiMlambda}) that $\bar{\hat{\Phi}}_{ABA'B'}o^{A'}
o^{B'}=0$ so that
$$
 \hat{\Phi}_{00'}=\hat{\Phi}_{01'}=\hat{\Phi}_{02'}=0.
$$
The `next' three components become, in GHP-formalism using Held's 
modified operators
\begin{eqnarray}
 \overline{\hat{\Phi}}_{10'} & = & \Ph L_{5}-(3\rho-\bar{\rho})L_{5}
 -\rho\Edt' L_{4}+2\Ph \ch_{01'}+2\bar{\rho}\ch_{01'} \nonumber \\
 \overline{\hat{\Phi}}_{11'} & = & \Phi_{11}+\Ph L_{6}-(2\rho-
 \bar{\rho})L_{6}-\rho\Edt' L_{5}+\Ph \ch_{11'}+(\rho+\bar{\rho})
 \ch_{11'}+\rho\Edt'\ch_{01'} \nonumber \\ \overline{\hat{\Phi}}_{12'}
 & = & \Phi_{21}+\Ph L_{7}-(\rho-\bar{\rho})L_{7}-\rho\Edt' L_{6}+2
 \rho\Edt'\ch_{11'} \label{midPhicomp}
\end{eqnarray}
We use (\ref{GHPWL}) and (\ref{GHPchL}) to eliminate the terms 
containing $\Edt'$ and use our expressions for the curvature 
components, Lanczos scalars and $\ch$-scalars to obtain the following
result from the first two equations of (\ref{midPhicomp}) 
$$
 \hat{\Phi}_{10'}=0\quad,\quad \hat{\Phi}_{11'}=0
$$
if and only if
\begin{eqnarray}
 \ch_{01'}^{\circ} & = & \frac{1}{6}\Edt' L_{4}^{\circ}+\frac{3}{2}
 \Omega^{\circ}L_{5}^{\circ} \nonumber \\ \ch_{11'}^{\circ} & = &
 -\frac{1}{2}\Edt' L_{5}^{\circ}-3\Omega^{\circ}\Lambda+\rho\bigl(\Edt'
 \ch_{01'}^{\circ}-\frac{1}{6}\Edt'^2 L_{4}^{\circ}-\frac{3}{2}L_{5}^{\circ}
 \Edt'\Omega^{\circ}-\frac{3}{2}\Omega^{\circ}\Edt' L_{5}^{\circ}\bigr).
 \label{}
\end{eqnarray}
respectively. We see that the expression within parenthesis vanishes
identically if $\hat{\Phi}_{10'}=0$ so that we obtain
$$
 \ch_{11'}^{\circ}=-\frac{1}{2}\Edt' L_{5}^{\circ}-3\Omega^{\circ}
 \Lambda.
$$
However, from (\ref{lambdaConditions}) we have that
$$
 \ch_{11'}^{\circ}=-\frac{1}{2}\Edt' L_{5}^{\circ}-\frac{1}{2}
 \Omega^{\circ}\Lambda
$$
so we obtain a necessary condition $\Omega^{\circ}\Lambda=0$.

Assuming the first two equations of (\ref{midPhicomp}) hold, the third
is easily seen to be equivalent to
\begin{equation}
	0=\Edt'^3 L_{4}^{\circ}+9\Omega^{\circ}\Edt'^2 L_{5}^{\circ}
	+9\Edt' \Omega^{\circ}\Edt' L_{5}^{\circ}+3L_{5}^{\circ}\Edt'^2
	\Omega^{\circ}+\Psi_{3}^{\circ},
	\label{}
\end{equation}
using our expressions for $\Phi_{21}$, $\Psi_{3}$ and $L_{7}$. This 
proves that
$$
 \hat{\Phi}_{10'}=\hat{\Phi}_{11'}=\hat{\Phi}_{12'}=0
$$
if and only if our class ${\cal G}$ spacetime with dyad in standard 
form is such that $\Lambda=0$ or 
$\Omega^{\circ}=0$ and in addition
\begin{eqnarray}
 \ch_{01'}^{\circ} & = & \frac{1}{6}\Edt' L_{4}^{\circ}+\frac{3}{2}
 \Omega^{\circ}L_{5}^{\circ} \nonumber \\ \ch_{11'}^{\circ} & = &
 -\frac{1}{2}\Edt' L_{5}^{\circ} \nonumber \\Ê0 & = & \Edt'^3
 L_{4}^{\circ}+9\Omega^{\circ}\Edt'^2 L_{5}^{\circ}+9\Edt' \Omega^{\circ}
 \Edt' L_{5}^{\circ}+3L_{5}^{\circ}\Edt'^2 \Omega^{\circ}+\Psi_{3}^{\circ}
 \label{PhimidConditions}
\end{eqnarray}
It remains to check that these choices satisfy the conditions from the 
previous chapters:
\begin{eqnarray}
 0 & = & \Edt'^3 L_{5}^{\circ}+\frac{1}{3}\Psi_{4}^{\circ} \nonumber \\
 0 & = & \Edt'^4 L_{4}^{\circ}+3L_{5}^{\circ}\Edt'^3\Omega^{\circ}+12
 \Edt'^2\Omega^{\circ}\Edt' L_{5}^{\circ}+18\Edt'\Omega^{\circ}\Edt'^2
 L_{5}^{\circ}+\Edt'\Psi_{3}^{\circ}-3\Omega^{\circ}\Psi_{4}^{\circ}
 \nonumber \\Ê0 & = & \Edt'^2\ch_{01'}^{\circ}+\ch_{11'}^{\circ}\Edt'
 \Omega^{\circ}-L_{5}^{\circ}\Edt'^2\Omega^{\circ}-\Edt'\Omega^{\circ}
 \Edt'L_{5}^{\circ}+\frac{1}{6}\Psi_{3}^{\circ} \label{prevCond}
\end{eqnarray}
We will now show that the equations (\ref{PhimidConditions}) and the 
first equation of (\ref{prevCond}) implies the last two equations 
of (\ref{prevCond}). First it is easily verified that the  second equation
of (\ref{prevCond}) can be rewritten
$$
 0=\Edt'\bigl(\Edt'^3 L_{4}^{\circ}+9\Omega^{\circ}\Edt'^2 L_{5}^{\circ}
 +9\Edt'\Omega^{\circ}\Edt' L_{5}^{\circ}+3L_{5}^{\circ}\Edt'^2
 \Omega^{\circ}+\Psi_{3}^{\circ}\bigr)-9\Omega^{\circ}\bigl(\Edt'^3
 L_{5}^{\circ}+\frac{1}{3}\Psi_{4}^{\circ}\bigr)
$$
so it is indeed identically satisfied. Substituting the first two 
equations of (\ref{PhimidConditions}) into the third equation of 
(\ref{prevCond}) it becomes, after simplification $\frac{1}{6}$ times 
the third equation of (\ref{PhimidConditions}) so it is also 
identically satisfied.

According to (\ref{HchiConditions}), any $H$-potential of the spinor 
$\Gamma_{ABCA'}$ must satisfy
\begin{eqnarray*}
 H_{12'}^{\circ} & = & 2\ch_{01'}^{\circ}-\frac{1}{3}\Edt'L_{4}^{\circ}=3
 \Omega^{\circ}L_{5}^{\circ} \nonumber \\ H_{22'}^{\circ} & = & 2
 \ch_{11'}^{\circ}-\frac{1}{2}\Edt' L_{5}^{\circ}=-\frac{3}{2}\Edt'
 L_{5}^{\circ}
\end{eqnarray*}
and in addition $H_{12'}^{\circ}$ must satisfy
\begin{eqnarray}
 0 & = & \Edt'^2 H_{12'}^{\circ}-3\Edt'\Omega^{\circ}\Edt'
 L_{5}^{\circ}+\frac{1}{3}\Edt'^3 L_{4}^{\circ}-2L_{5}^{\circ}\Edt'^2
 \Omega^{\circ}+\frac{1}{3}\Psi_{3}^{\circ} \nonumber \\ & = & \frac{1}{3}
 \bigl(\Edt'^3 L_{4}^{\circ}+9\Omega^{\circ}\Edt'^2 L_{5}^{\circ}+9
 \Edt'\Omega^{\circ}\Edt' L_{5}^{\circ}+3L_{5}^{\circ}\Edt'^2
 \Omega^{\circ}+\Psi_{3}^{\circ}\bigr)
\end{eqnarray}
which is identically satisfied. This proves the following result,

\begin{lemma}
 Given a spacetime of class ${\cal G}$ with dyad in standard form, there
 exists a spinor $H_{ABA'B'}=Q_{AB}o_{A'}o_{B'}$, 
 $Q_{AB}=Q_{(AB)}$ such that the spinor
 $$
  \Gamma_{ABCA'}=\nabla_{(A}{}^{B'}H_{B)CA'B'}
 $$
 defines a metric, asymmetric connection for which all curvature quantities
 vanish except $\hat{\Phi}_{20'}$, $\hat{\Phi}_{21'}$ and $\hat{\Phi}_{22'}$
 if and only if $\Lambda=0$ or $\Omega^{\circ}=0$ 
 $(\Leftrightarrow \rho=\bar{\rho})$. All such spinors $H_{ABA'B'}$ are 
 given by
 \begin{eqnarray}
  H_{02'} & = & \frac{\bar{\rho}}{\rho}L_{4}^{\circ}+\bar{\rho}H_{02'}^{\circ}
  \nonumber \\ H_{12'} & = & \frac{3}{2}\frac{\bar{\rho}}{\rho^2}L_{5}^{\circ}
  +3\frac{\bar{\rho}}{\rho}\Omega^{\circ}L_{5}^{\circ}-\frac{1}{2}\bar{\rho}
  \bigl(\Edt' H_{02'}^{\circ}-L_{4}^{\circ}\Edt' \Omega^{\circ}\bigr)
  \nonumber \\ H_{22'} & = & -3\frac{\bar{\rho}}{\rho^3}\Lambda-\frac{3}{2}
  \frac{\bar{\rho}}{\rho^2}\Edt' L_{5}^{\circ}-\frac{1}{2}\frac{\bar{\rho}}
  {\rho}\bigl(\Edt'^2 L_{4}^{\circ}+12\Omega^{\circ}\Edt' L_{5}^{\circ}+3
  L_{5}^{\circ}\Edt'\Omega^{\circ}\bigr)+\frac{1}{4}\rho\Psi_{2}^{\circ}
  \nonumber \\ & & +\frac{1}{2}\bar{\rho}\bigl(\Edt'^2 H_{02'}^{\circ}+6
  L_{5}^{\circ}\Omega^{\circ}\Edt'\Omega^{\circ}-L_{4}^{\circ}\Edt'^2
  \Omega^{\circ}-\Edt' \Omega^{\circ}\Edt' L_{4}^{\circ}+\frac{1}{2}
  \Psi_{2}^{\circ}\bigr) \nonumber \\ & & +\frac{1}{2}\rho\bar{\rho}
  \Phi_{11}^{\circ}
  \label{allHpot3}
 \end{eqnarray}
 where $L_{4}^{\circ}, L_{5}^{\circ}$ and $H_{02'}^{\circ}$ are subject to
 the conditions
 \begin{eqnarray}
   0 & = & \Edt'^3 L_{5}^{\circ}+\frac{1}{3}\Psi_{4}^{\circ} \nonumber
   \\ 0 & = & \Edt'^3 L_{4}^{\circ}+3L_{5}^{\circ}\Edt'^2 \Omega^{\circ}
   +9\Edt'\Omega^{\circ}\Edt' L_{5}^{\circ}+9\Omega^{\circ}\Edt'^2
   L_{5}^{\circ}+\Psi_{3}^{\circ} \nonumber \\ 0 & = & \Edt'^3
   H_{02'}^{\circ}+6L_{5}^{\circ}\Omega^{\circ}\Edt'^2\Omega^{\circ}
   +18\Omega^{\circ}\Edt'\Omega^{\circ}\Edt' L_{5}^{\circ}-L_{4}^{\circ}
   \Edt'^3\Omega^{\circ}-2\Edt'^2 \Omega^{\circ}\Edt' L_{4}^{\circ} \nonumber
   \\ & & +9L_{5}^{\circ}(\Edt' \Omega^{\circ})^2+\frac{1}{2}\Edt'
   \Psi_{2}^{\circ}-\Omega^{\circ}\Psi_{3}^{\circ}-\Phi_{21}^{\circ} 
   \label{LHConditions3}
 \end{eqnarray}
\end{lemma}

The remaining components of equation (\ref{PhiMlambda}) can now be 
written in GHP-formalism, using Held's modified operators, as
\begin{eqnarray}
 \overline{\hat{\Phi}}_{20'} & = & -2\Pht' L_{4}+\bigl[2\rho'+
 \frac{3\Psi_{2}}{\rho}-\frac{\bar{\Psi}_{2}}{\bar{\rho}}+\frac{4\Lambda}
 {\rho}\bigr]L_{4}+2\bar{\rho}\Edt L_{5}+4\bar{\rho}\Edt\ch_{01'} \nonumber
 \\ & & +8\bigl(L_{6}L_{4}-L_{5}^2+L_{4}\ch_{11'}-L_{5}\ch_{01'}+2\ch_{01'}^2
 \bigr) \nonumber \\ \overline{\hat{\Phi}}_{21'} & = & \Phi_{12}-2\Pht'
 L_{5}+\bigl[4\rho'+\frac{\Psi_{2}}{\rho}-\frac{\bar{\Psi}_{2}}{\bar{\rho}}
 \bigr]L_{5}+2\bar{\rho}\Edt L_{6}-2\kappa' L_{4} \nonumber \\Ê& & +2\Pht'
 \ch_{01'}+\bigl[2\rho'-\frac{\Psi_{2}}{\rho}+\frac{\bar{\Psi}_{2}}{\bar{\rho}}
 \bigr]\ch_{01'}+2\bar{\rho}\Edt\ch_{11'} \nonumber \\Ê& & +4\bigl(L_{4}L_{7}
 -L_{5}L_{6}+2L_{5}\ch_{11'}-2L_{6}\ch_{01'}+4\ch_{01'}\ch_{11'}\bigr)
 \nonumber \\ \overline{\hat{\Phi}}_{22'} & = & \Phi_{22}-2\Pht' L_{6}
 +\bigl[6\rho'-\frac{\Psi_{2}}{\rho}-\frac{\bar{\Psi}_{2}}{\bar{\rho}}
 -\frac{4\Lambda}{\rho}\bigr]L_{6}+2\bar{\rho}\Edt L_{7}-4\kappa' L_{5}
 \nonumber \\ & & +4\Pht'\ch_{11'}+\bigl[\frac{2\Psi_{2}}{\rho}+
 \frac{2\bar{\Psi}_{2}}{\bar{\rho}}+\frac{8\Lambda}{\rho}\bigr]\ch_{11'}
 +4\kappa'\ch_{01'} \nonumber \\Ê& & +8\bigl(L_{5}L_{7}-L_{6}^2+L_{6}
 \ch_{11'}-L_{7}\ch_{01'}+2\ch_{11'}^2\bigr) \label{nonlinPhis}
\end{eqnarray}
where we have used that $\Omega^{\circ}\Lambda=0$ and hence that 
$\frac{\Lambda}{\rho}=\frac{\Lambda}{\bar{\rho}}$. At a first glance 
it seems unlikely that these equations can be solved since they are 
highly non-linear, but we shall see that the situation is manageable.
We will start by looking at the non-twisting case, i.e., $\Omega^{\circ}=0$
(it is not necessary to make this separation into two cases $\Omega^{\circ}
=0$ and $\Omega^{\circ}\neq 0$, but it simplifies the calculations 
greatly). By substituting our previous equations into the first equation
of (\ref{nonlinPhis}) we find that $\hat{\Phi}_{20'}=0$ if and only if
\begin{eqnarray}
 0 & = & \Lambda L_{4}^{\circ} \nonumber \\ 0 & = & 3\Edt L_{5}^{\circ}-\Pht'
 L_{4}^{\circ}-6L_{4}^{\circ}\Edt'L_{5}^{\circ}+6L_{5}^{\circ}\Edt'
 L_{4}^{\circ}
\end{eqnarray}
Continuing with the second equation of (\ref{nonlinPhis}) we obtain 
that $\hat{\Phi}_{20'}=\hat{\Phi}_{21'}=0$ if and only if
\begin{eqnarray}
 0 & = & \Lambda L_{4}^{\circ} \nonumber \\ 0 & = & \Lambda L_{5}^{\circ}
 \nonumber \\ 0 & = & 3\Edt L_{5}^{\circ}-\Pht'L_{4}^{\circ}-6L_{4}^{\circ}
 \Edt'L_{5}^{\circ}+6L_{5}^{\circ}\Edt'L_{4}^{\circ}
\end{eqnarray}
After a very long calculation the last equation of (\ref{nonlinPhis}) 
gives us that $\hat{\Phi}_{20'}=\hat{\Phi}_{21'}=\hat{\Phi}_{22'}=0$,
and hence that $\hat{R}_{abcd}=0$ if and only if $\Lambda=0$ and in 
addition
\begin{equation}
 3\Edt L_{5}^{\circ}-\Pht'L_{4}^{\circ}-6L_{4}^{\circ}\Edt'L_{5}^{\circ}+
 6L_{5}^{\circ}\Edt'L_{4}^{\circ}=0
	\label{}
\end{equation}
along with all other conditions derived previously. Before we look at 
the possibility of satisfying all the conditions we have obtained, we 
will also look at the non-twisting case $\Omega^{\circ}\neq 0$. Then, 
by our previous conditions we already have $\Lambda=0$. If we 
substitute our previous equations into the three equations 
(\ref{nonlinPhis}), a very long calculation indeed reveals that
$\hat{\Phi}_{20'}=\hat{\Phi}_{21'}=\hat{\Phi}_{22'}=0$,
and hence that $\hat{R}_{abcd}=0$ if and only if
\begin{equation}
 3\Edt L_{5}^{\circ}-\Pht'L_{4}^{\circ}-6L_{4}^{\circ}\Edt'L_{5}^{\circ}+
 6L_{5}^{\circ}\Edt'L_{4}^{\circ}+18\Omega^{\circ}L_{5}^{\circ 2}=0
	\label{}
\end{equation}
along with the previously derived conditions.

This proves the following result
\begin{theorem}
 In a spacetime of class ${\cal G}$ with dyad in standard form a necessary
 condition for $L_{ABCA'}$ and $\ch_{AA'}$ to define a completely
 curvature-free connection is that $\Lambda=0$. All such connections are
 given by (\ref{Lscalars}) and (\ref{chiscalars}) where the functions of 
 integration satisfy the conditions
 \begin{eqnarray}
  \ch_{01'}^{\circ} & = & \frac{1}{6}\Edt' L_{4}^{\circ}+\frac{3}{2}
  \Omega^{\circ}L_{5}^{\circ} \nonumber \\ \ch_{11'}^{\circ} & = &
  -\frac{1}{2}\Edt' L_{5}^{\circ} \nonumber \\Ê0 & = & L_{7}^{\circ} 
  \nonumber \\ 0 & = & L_{6}^{\circ} \nonumber \\ 0 & = & \Edt'^3 L_{5}^{\circ}
  +\frac{1}{3}\Psi_{4}^{\circ} \nonumber \\ 0 & = & \Edt'^3 L_{4}^{\circ}+9
  \Omega^{\circ}\Edt'^2 L_{5}^{\circ}+9\Edt' \Omega^{\circ}\Edt'
  L_{5}^{\circ}+3L_{5}^{\circ}\Edt'^2 \Omega^{\circ}+\Psi_{3}^{\circ}
  \nonumber \\ 0 & = & 3\Edt L_{5}^{\circ}-\Pht'L_{4}^{\circ}-6L_{4}^{\circ}
  \Edt'L_{5}^{\circ}+6L_{5}^{\circ}\Edt'L_{4}^{\circ}+18\Omega^{\circ}
  L_{5}^{\circ 2} \label{necsuffcond}
 \end{eqnarray}
 All $H$-potentials satisfying $\nabla_{(A}{}^{B'}H_{B)CA'B'}=\Gamma_{ABCA'}$
 are given by equation (\ref{allHpot3}) subject to the condition
 \begin{eqnarray}
  0 & = & \Edt'^3 H_{02'}^{\circ}+6L_{5}^{\circ}\Omega^{\circ}\Edt'^2
  \Omega^{\circ}+18\Omega^{\circ}\Edt'\Omega^{\circ}\Edt' L_{5}^{\circ}-
  L_{4}^{\circ}\Edt'^3\Omega^{\circ}-2\Edt'^2 \Omega^{\circ}\Edt' L_{4}^{\circ}
  \nonumber \\ & & +9L_{5}^{\circ}(\Edt' \Omega^{\circ})^2+\frac{1}{2}\Edt'
   \Psi_{2}^{\circ}-\Omega^{\circ}\Psi_{3}^{\circ}-\Phi_{21}^{\circ}
 \end{eqnarray}
\end{theorem}
Note that at this moment we have not yet proved that all these 
conditions can be simultaneously satisfied.

\subsubsection{The existence of completely curvature-free connections}

In this section we will show that $\Lambda=0$ is also a sufficient 
condition for the existence of a curvature-free connection of the type 
discussed previously. As seen in the previous theorem we need to find 
a solution to
the equations
\begin{eqnarray}
 0 & = & \Edt'^3 L_{5}^{\circ}+\frac{1}{3}\Psi_{4}^{\circ} \nonumber \\ 0
 & = & \Edt'^3 L_{4}^{\circ}+9\Omega^{\circ}\Edt'^2 L_{5}^{\circ}+9\Edt'
 \Omega^{\circ}\Edt'L_{5}^{\circ}+3L_{5}^{\circ}\Edt'^2 \Omega^{\circ}+
 \Psi_{3}^{\circ} \nonumber \\ 0 & = & 3\Edt L_{5}^{\circ}-\Pht'L_{4}^{\circ}
 -6L_{4}^{\circ}\Edt'L_{5}^{\circ}+6L_{5}^{\circ}\Edt'L_{4}^{\circ}+18
 \Omega^{\circ}L_{5}^{\circ 2} \label{curvfreecond}
\end{eqnarray}
We observe that the first equation can be written
$$
 0=\Edt'^3 L_{5}^{\circ}+\frac{1}{3}\Psi_{4}^{\circ}=\Edt'\bigl(\Edt'^2
 L_{5}^{\circ}-\frac{1}{3}\kappa'^{\circ}\bigr)
$$
Thus, the first equation is satisfied, e.g., if
\begin{equation}
	\Edt'^2 L_{5}^{\circ}=\frac{1}{3}\kappa'^{\circ}.
	\label{}
\end{equation}
We observe that via the commutators it is easy to show that there 
actually exists functions $L_{5}^{\circ}$ that satisfy this 
equation. Then the second equation of (\ref{curvfreecond}) can be
rewritten
$$
 0=\Edt'\bigl(\Edt'^2 L_{4}^{\circ}+3L_{5}^{\circ}\Edt'\Omega^{\circ}
 +6\Omega^{\circ}\Edt'L_{5}^{\circ}-\rho'^{\circ}\bigr)
$$
so it in turn is satisfied if, e.g.,
\begin{equation}
	\Edt'^2 L_{4}^{\circ}=\rho'^{\circ}-3L_{5}^{\circ}\Edt'\Omega^{\circ}
    -6\Omega^{\circ}\Edt'L_{5}^{\circ}.
	\label{}
\end{equation}
We note that $L_{4}^{\circ}$ satisfies this equation if and only if it
also satisfies the condition
\begin{equation}
	\Edt'L_{4}^{\circ}=-3\Omega^{\circ}L_{5}^{\circ}+\alpha^{\circ}
	\label{}
\end{equation}
for some function $\alpha^{\circ}$ that satisfies
\begin{equation}
	\Edt'\alpha^{\circ}=\rho'^{\circ}-3\Omega^{\circ}\Edt'L_{5}^{\circ}.
	\label{}
\end{equation}
Applying the $\Bigl[\Pht',\Edt'\Bigr]$-commutator to $L_{4}^{\circ}$ 
then gives us the following necessary and sufficient condition for 
the existence of a solution $L_{4}^{\circ}$:
$$
 \Pht'\alpha^{\circ}=3\Edt\Edt'L_{5}^{\circ}.
$$
Applying the same commutator to $\alpha^{\circ}$ we find that it is 
identically satisfied and hence there exists a function $\alpha^{\circ}$
that satisfies both of the above conditions. It follows that the 
conditions for $L_{4}^{\circ}$ also satisfies the commutators 
identically, and therefore there actually exists solutions of 
(\ref{curvfreecond}).

Thus, our final result is
\begin{theorem}
 In a spacetime of class ${\cal G}$ with dyad in standard form there exists
 a Lanczos potential of the Weyl spinor $L_{ABCA'}$ and a covector $\ch_{AA'}$,
 both aligned to $o^{A'}$ such that the resulting connection
 $\hat{\nabla}_{AA'}$ is completely curvature-free (i.e., $\hat{R}_{abcd}=0$)
 if and only if $\Lambda=0$.
 
 A possible choice of $L_{ABCA'}$ and $\ch_{AA'}$ is given by
 \begin{eqnarray}
  L_{4} & = & \bar{\rho}L_{4}^{\circ} \nonumber \\ L_{5} & = &
  \frac{\bar{\rho}}{\rho}L_{5}^{\circ}-\frac{1}{3}\bar{\rho}\Edt'
  L_{4}^{\circ} \nonumber \\ L_{6} & = & -\frac{\bar{\rho}}{\rho}\Edt'
  L_{5}^{\circ}+\frac{1}{6}\bar{\rho}\bigl(\rho'^{\circ}-6\Omega^{\circ}
  \Edt'L_{5}^{\circ}\bigr)-\frac{1}{4}\rho^2\Psi_{2}^{\circ}-\frac{1}{12}
  \rho\bar{\rho}\Psi_{2}^{\circ}-\frac{1}{2}\rho^2\bar{\rho}\Phi_{11}^{\circ} 
  \nonumber \\ L_{7} & = & \frac{1}{2}\kappa'^{\circ}-\frac{1}{2}\rho
  \Psi_{3}^{\circ}-\frac{1}{4}\rho^2 \Edt'\Psi_{2}^{\circ}-\frac{1}{2}\rho
  \bar{\rho}\Phi_{21}^{\circ}-\frac{1}{4}\rho^3\Psi_{2}^{\circ}\Edt'
  \Omega^{\circ}-\frac{1}{2}\rho^2\bar{\rho}\Edt'\Phi_{11}^{\circ}
  \nonumber \\ & & -\frac{1}{2}\rho^3\bar{\rho}\Phi_{11}^{\circ}\Edt'
  \Omega^{\circ} \nonumber \\ \ch_{01'} & = & \frac{\bar{\rho}}{\rho}
  L_{5}^{\circ}+\bar{\rho}\bigl(\frac{1}{6}\Edt'L_{4}^{\circ}+\frac{3}{2}
  \Omega^{\circ}L_{5}^{\circ}\bigr) \nonumber \\ \ch_{11'} & = & 
  -\frac{1}{2}\Edt'L_{5}^{\circ}-\frac{1}{6}\bar{\rho}\rho'^{\circ}
  +\frac{1}{12}\rho\bar{\rho}\Psi_{2}^{\circ}
  \label{LscalarsFlat}
 \end{eqnarray}
 where
 \begin{eqnarray}
  \Edt'^2 L_{5}^{\circ} & = & \frac{1}{3}\kappa'^{\circ} \nonumber \\ \Edt'^2
  L_{4}^{\circ} & = & \rho'^{\circ}-3L_{5}^{\circ}\Edt'\Omega^{\circ}-6
  \Omega^{\circ}\Edt'L_{5}^{\circ} \nonumber \\ \Pht'L_{4}^{\circ} & = &
  3\Edt L_{5}^{\circ}-6L_{4}^{\circ}\Edt'L_{5}^{\circ}+6L_{5}^{\circ}\Edt'
  L_{4}^{\circ}+18\Omega^{\circ}L_{5}^{\circ 2}
 \end{eqnarray}
 and in particular, there always exists functions $L_{4}^{\circ}$, 
 $L_{5}^{\circ}$ satisfying these conditions.
 
 All $H$-potentials of these connections that are aligned to $o^{A'}$ 
 are given by
 \begin{eqnarray}
  H_{02'} & = & \frac{\bar{\rho}}{\rho}L_{4}^{\circ}+\bar{\rho}H_{02'}^{\circ}
  \nonumber \\ H_{12'} & = & -\frac{3}{2}\frac{\bar{\rho}}{\rho^2}L_{5}^{\circ}
  +\frac{3}{\rho}L_{5}^{\circ}-\frac{1}{2}\bar{\rho}\bigl(\Edt'
  H_{02'}^{\circ}-L_{4}^{\circ}\Edt' \Omega^{\circ}\bigr) \nonumber \\
  H_{22'} & = & \frac{3}{2}\frac{\bar{\rho}}{\rho^2}\Edt'L_{5}^{\circ}
  -\frac{3}{\rho}\Edt'L_{5}^{\circ}-\frac{1}{2}\frac{\bar{\rho}}{\rho}
  \rho'^{\circ}+\frac{1}{4}\rho \Psi_{2}^{\circ}+\frac{1}{2}\bar{\rho}\bigl(
  \Edt'^2 H_{02'}^{\circ} \nonumber \\ & & +2H_{12'}^{\circ}\Edt'\Omega^{\circ}
  -L_{4}^{\circ}\Edt'^2 \Omega^{\circ}-\Edt' \Omega^{\circ}\Edt' L_{4}^{\circ}
  +\frac{1}{2}\Psi_{2}^{\circ}\bigr)+\frac{1}{2}\rho\bar{\rho}\Phi_{11}^{\circ}
  \label{allHpotLflat}
 \end{eqnarray}
 where $H_{02'}^{\circ}$ satisfies
 \begin{eqnarray}
  0 & = & \Edt'^3 H_{02'}^{\circ}+6L_{5}^{\circ}\Omega^{\circ}\Edt'^2
  \Omega^{\circ}+18\Omega^{\circ}\Edt'\Omega^{\circ}\Edt' L_{5}^{\circ}-
  L_{4}^{\circ}\Edt'^3\Omega^{\circ}-2\Edt'^2 \Omega^{\circ}\Edt' L_{4}^{\circ}
  \nonumber \\ & & +9L_{5}^{\circ}(\Edt' \Omega^{\circ})^2+\frac{1}{2}\Edt'
   \Psi_{2}^{\circ}-\Omega^{\circ}\Psi_{3}^{\circ}-\Phi_{21}^{\circ}
 \end{eqnarray}
 and in particular, such a function $H_{02'}^{\circ}$ exists.
\end{theorem}

\section{Applications to quasi-local momentum}

\subsection{Quasi-local momentum in spacetimes of class ${\cal G}$}

Now that we have obtained curvature-free connections in the spacetimes 
of class ${\cal G}$, we will look at possible applications to physics.
Thus, in this section we will see how far the Bergqvist-Ludvigsen construction
of quasi-local momentum can be taken in a general class ${\cal G}$ 
spacetime. In an analogous way as for the Kerr spacetime, let ${\cal S}_{A}$
denote the 2-dimensional complex vector space of spinor fields $\xi_{A}$
satisfying
\begin{equation}
	\hat{\nabla}_{AA'}\xi_{B}=0.
	\label{xiinS}
\end{equation}
where $\hat{\nabla}_{AA'}$ is an arbitrary curvature-free connection 
given in Theorem 4.5. Put
\begin{equation}
	\varphi_{AB}=\xi_{(A}\nabla_{B)}{}^{C'}\bar{\xi}_{C'}-\bar{\xi}_{C'}
	\nabla_{(A}{}^{C'}\xi_{B)} \label{phidef}
\end{equation}
and
\begin{equation}
	F_{ab}=i\bigl(\eps_{AB}\bar{\varphi}_{A'B'}-\eps_{A'B'}\varphi_{AB}
	\bigr). \label{NW}
\end{equation}
Given a spacelike 2-surface $\Sigma$ we now define a 1-form $P_{AA'}$ on
the hermitian part of ${\cal S}^{A}\otimes\bar{{\cal S}}^{A'}$ by
\begin{equation}
	P_{AA'}(\Sigma)\xi^{A}\bar{\xi}^{A'}=\frac{1}{8\pi}\int_{\Sigma}
	{\bf F},
	\label{}
\end{equation}
analogously to \cite{BL}, \cite{BL1}.

Because $F_{ab}$ is a 2-form, $(dF)_{abc}=\nabla_{[a}F_{bc]}$ is a 
3-form so its Hodge dual $({}^{*}dF)_{a}$ is a 1-form which is much 
easier to calculate than $(dF)_{abc}$ and we have that
$$
 ({}^{*}dF)_{a}=\nabla_{A'}{}^{B}\varphi_{AB}+\nabla_{A}{}^{B'}
 \bar{\varphi}_{A'B'}.
$$
By using (\ref{xiinS}) we obtain
\begin{equation}
 \varphi_{AB}=2\bigl(\bar{\Gamma}_{C'D'}{}^{D'}{}_{(A}\eps_{B)C}-
 \Gamma_{C(AB)C'}\bigr)\xi^{C}\bar{\xi}^{C'}.
\end{equation}
Decomposing $\Gamma_{ABCA'}$ yields 
\begin{equation}
 \varphi_{AB}=2\bigl(3\bar{\ch}_{C'(A}\eps_{B)C}+\ch_{C'(A}\eps_{B)C}
 -L_{ABCC'}\bigr)\xi^{C}\bar{\xi}^{C'}.
\end{equation}
A very long spinor calculation involving both the equations (\ref{DecompRhat})
and (\ref{xiinS}) now reveals that
\begin{eqnarray}
	({}^{*}dF)_{a} & = & -2\xi^{B}\bar{\xi}^{B'}\bigl(\Phi_{ABA'B'}+4(M_{ABC}
	o^{C}-\lambda_{A}o_{B}+2o_{A}\lambda_{B}) \nonumber \\ & & \cdot(
	\bar{M}_{A'B'C'}o^{C'}-\bar{\lambda}_{A'}o_{B'}+2o_{A'}\bar{\lambda}_{B'}
	)-36o_{A}o_{A'}\lambda_{B}\bar{\lambda}_{B'}\bigr) \nonumber \\
	& =: & -\xi^{B}\bar{\xi}^{B'}\bigl(\Phi_{ABA'B'}+{\cal F}_{ABA'B'}+
	\eps_{A'B'}{\cal E}_{AB}+\eps_{AB}\bar{{\cal E}}_{A'B'}\bigr)
	\label{}
\end{eqnarray}
where $L_{ABCA'}=M_{ABC}o_{A'}$ and $\chi_{AA'}=\lambda_{A}o_{A'}$.
Explicitly, the hermitian spinor ${\cal F}_{ABA'B'}={\cal F}_{(AB)(A'B')}$
and the spinor ${\cal E}_{AB}={\cal E}_{(AB)}$ are given by
\begin{eqnarray}
	{\cal F}_{ABA'B'} & = & 4(M_{ABC}o^{C}+o_{(A}\lambda_{B)})
	(\bar{M}_{A'B'C'}o^{C'}+o_{(A'}\bar{\lambda}_{B')}) \nonumber \\ & &
	-36o_{(A}\lambda_{B)}o_{(A'}\bar{\lambda}_{B')}\bigr) \nonumber \\
	{\cal E}_{AB} & = & 6o_{A'}\bar{\lambda}^{A'}(M_{ABC}o^{C}-2o_{(A}
	\lambda_{B)}). \label{dFeqn}
\end{eqnarray}
The components of ${\cal F}_{ABA'B'}$ and ${\cal E}_{AB}$ in a spinor dyad
$(o^{A},\iota^{A})$ with $\iota^{A}$ arbitrary, are given by
\begin{eqnarray}
	{\cal E}_{0} & = & -6\overline{\ch_{01'}}L_{4} \nonumber \\
	{\cal E}_{1} & = & -6\overline{\ch_{01'}}(\ch_{01'}-L_{5}) \nonumber \\
	{\cal E}_{2} & = & -6\overline{\ch_{01'}}(L_{6}-2\ch_{11'}) \nonumber \\
	{\cal F}_{00'} & = & 4L_{4}\overline{L_{4}} \nonumber \\
	{\cal F}_{10'} & = & 2\overline{L_{4}}(2L_{5}+\ch_{01'}) \nonumber \\
	{\cal F}_{20'} & = & 4\overline{L_{4}}(L_{6}+\ch_{11'}) \nonumber \\
	{\cal F}_{11'} & = & (2L_{5}+\ch_{01'})(2\overline{L_{5}}+
	\overline{\ch_{01'}})-9\ch_{01'}\overline{\ch_{01'}} \nonumber \\
	{\cal F}_{21'} & = & 2(L_{6}+\ch_{11'})(2\overline{L_{5}}+
	\overline{\ch_{01'}})-18\ch_{11'}\overline{\ch_{01'}} \nonumber \\
	{\cal F}_{22'} & = & 4(L_{6}+\ch_{11'})(\overline{L_{6}}+
	\overline{\ch_{11'}})-36\ch_{11'}\overline{\ch_{11'}}
	\label{dFcompeqn}
\end{eqnarray}
We remark that in an asymptotically flat spacetime an analogous 
construction can be performed. As our spin space ${\cal S}$ we take 
the asymptotic spin space \cite{PR2}. For $\xi^{A}$ asymptotically 
constant we define $\varphi_{AB}$ as in (\ref{phidef}) and $F_{ab}$
as in (\ref{NW}). Then $F_{ab}$ is called the Nester-Witten 2-form, the 
resulting momentum $P_{AA'}(\Sigma_{\infty})$ where $\Sigma_{\infty}$ 
is a spacelike cross-section of future null infinity, is called the Bondi
momentum and the Hodge dual of the 1-form $-\xi^{B}\bar{\xi}^{B'}\bigl(
{\cal F}_{ABA'B'}+\eps_{A'B'}{\cal E}_{AB}+\eps_{AB}\bar{{\cal E}}_{A'B'}
\bigr)$ is called the Sparling 3-form \cite{PR2}.

We recall that in the Bergqvist-Ludvigsen construction, $F_{ab}$ was a 
closed 2-form. For $F_{ab}$ to be closed in the more general class ${\cal G}$
vacuum case it is necessary (\ref{dFeqn}) that $\ch_{01'}=\lambda_{A}o^{A}=0$
or that $M_{ABC}o^{C}=2o_{(A}\lambda_{B)}$.

We first consider the case $M_{ABC}o^{C}=2o_{(A}\lambda_{B)}$, i.e., in
components $L_{4}=0$, $L_{5}=\ch_{01'}$ and $L_{6}=2\ch_{11'}$ (from
(\ref{dFcompeqn})). In a spinor dyad in standard form, the functions of
integration must satisfy $L_{4}^{\circ}=0$, $\ch_{01'}^{\circ}=0$,
$L_{5}^{\circ}\Edt'\Omega^{\circ}=0$ and also $\Psi_{2}^{\circ}=0$ according
to the equations (\ref{Lscalars}), (\ref{chiscalars}) and (\ref{necsuffcond}).
This implies that $\Psi_{2}=0$ so the spacetime has to be at least Petrov type
III. Thus, the condition $M_{ABC}o^{C}=2o_{(A}\lambda_{B)}$ places severe
restrictions on a vacuum spacetime.

We also see that if $M_{ABC}o^{C}\neq 2o_{(A}\lambda_{B)}$, the only other
possibility for $F_{ab}$ to be closed is that $\ch_{01'}=0$. In this case we
also obtain $L_{4}=0$, $L_{5}=0$ and in addition
$$
 (L_{6}+\ch_{11'})(\overline{L_{6}}+\overline{\ch_{11'}})=9\ch_{11'}
 \overline{\ch_{11'}}.
$$
Referring to (\ref{Lscalars}) and (\ref{chiscalars}) we find that the 
functions of integration must satisfy $L_{4}^{\circ}=0$, $L_{5}^{\circ}=0$
and $\ch_{01'}^{\circ}=0$. These are also very restrictive conditions even 
though the last one is seen to be identically satisfied. From 
(\ref{necsuffcond}) we see that the vacuum spacetime must satisfy 
$\Psi_{3}^{\circ}=0$ and $\Psi_{4}^{\circ}=0$.

\subsection{Kerr-Schild spacetimes of class ${\cal G}$ with vanishing 
Ricci scalar}

As an application of the results in the previous section we will now 
look at Kerr-Schild spacetimes of class ${\cal G}$ with vanishing 
Ricci scalar. Following the conventions of Section 4.2 we obtain the
Lanczos- and $\ch$-scalars
\begin{eqnarray}
	L_{4} & = & 0\quad,\quad L_{5}=0\quad,\quad \ch_{01'}=0 \nonumber \\
	L_{6} & = & -\frac{1}{6}\bigl(\Ph f+(2\rho-\bar{\rho})f\bigr) 
	\nonumber \\ L_{7} & = & -\frac{1}{2}\bigl(\Ed'f-\bar{\tau}f\bigr)
	\nonumber \\ \ch_{11'} & = & -\frac{1}{12}\bigl(\Ph f-(\rho+
	\bar{\rho})f\bigr)
	\label{}
\end{eqnarray}
for arbitrary dyad spinor $\iota^{A}$, so we allow for the possibility of the
dyad not being in standard form. Then we immediately obtain ${\cal E}_{AB}=0$
and in addition
\begin{eqnarray}
	{\cal F}_{00'} & = & 0\quad,\quad{\cal F}_{10'}=0\quad,\quad
	{\cal F}_{20'}=0 \nonumber \\ {\cal F}_{11'} & = & 0
	\quad,\quad {\cal F}_{21'}=0 \nonumber \\ {\cal F}_{22'}
	& = & \frac{f}{2}\bigl((\rho+\bar{\rho})\Ph f-(\rho^2+
	\bar{\rho}^2)f\bigr).
	\label{}
\end{eqnarray}
However, it is easily shown that in these spacetimes
$$
 (\rho+\bar{\rho})\Ph f-(\rho^2+\bar{\rho}^2)f=-2\Phi_{11}
$$
by rewriting the relevant Newman-Penrose equations in \cite{KSMH}. Hence,
$$
 {\cal F}_{22'}=-f\Phi_{11}
$$
and we can therefore write
\begin{equation}
	{\cal F}_{ABA'B'}=-f\Phi_{11}o_{A}o_{B}o_{A'}o_{B'}.
	\label{}
\end{equation}
We see that in particular the 2-form $F_{ab}$ is closed if and only 
if the Kerr-Schild spacetime is vacuum, similarly to the 
Bergqvist-Ludvigsen construction in the Kerr spacetime. Hence, if 
$\Sigma_{1}$ and $\Sigma_{2}$ are two spacelike hypersurfaces 
such that they together form the boundary of some 3-volume $V$, then
$P_{AA'}(\Sigma_{1})=P_{AA'}(\Sigma_{2})$ according to Stokes' 
theorem, in the vacuum case.

\section{Conclusions}

In spacetimes of class ${\cal G}$ with dyad in standard form we obtained,
by the method of $\rho$-integration, all Lanczos potentials that are
aligned to $o^{A'}$, of the Weyl spinor and their $H$-potentials (also 
aligned to $o^{A'}$). The resulting expressions for the Lanczos scalars can
be written as polynomials in $\rho$ and $\rho^{-1}$, divided by some power
of the factor $(1+\rho\Omega^{\circ})$, by making use of the formula
$$
 \bar{\rho}=\frac{\rho}{1+\rho\Omega^{\circ}}.
$$
This is closely related to the peeling theorem in asymptotically flat
spacetimes. We therefore expect it to be possible to, extend the approach
in this paper to such spacetimes and so it may be possible to integrate
for Lanczos potentials and use them to construct curvature-free connections
for (some) asymptotically flat spacetimes.

We remark that this paper can be viewed as an alternative existence
proof for Lanczos potentials of the Weyl spinor and $H$-potentials of
Lanczos potentials of the Weyl spinor, for spacetimes of class ${\cal 
G}$. We also remark that the existence proof for $H$-potentials of a 
general symmetric (3,1)-spinor in \cite{AE4} is valid only in Einstein
spacetimes, whereas we have found $H$-potentials in the special case 
when $L_{ABCA'}$ is a Lanczos potential of the Weyl spinor that is 
aligned to the repeated principal spinor. A similar existence proof was
obtained by Torres del Castillo \cite{TdC1}, \cite{TdC2} for a slightly
more general class of spacetimes though, as mentioned above, he did not
find all potentials of the type we have discussed. His approach was
reminiscent of the ${\cal H}$-space theory \cite{KLNT}; it would be
interesting to investigate which of the potentials found in this paper
can be written in the form that he derived.

We also note that the condition that $L_{ABCA'}$ possesses an 
$H$-potential aligned to $o^{A'}$ is actually a {\em necessary} 
condition for $\Gamma_{ABCA'}$ to define a curvature-free connection 
in the case that we have studied (Theorem \ref{Sigmaflat}). This is an 
interesting result and it raises the question whether $H$-potentials 
of Lanczos potentials of the Weyl spinor offers possibilities for 
constructing curvature-free connections and quasi-local momentum in 
more general spacetimes. We also remark that hermitian $H$-potentials 
seem to play a role in the construction of angular momentum \cite{BL1}. 
It would therefore be of interest to investigate when hermitian
$H$-potentials can be found.

It has been conjectured \cite{ACL} that the Lanczos potential is related to
the NP spin coefficients. In \cite{AE3} Lanczos potentials for the Weyl
spinor whose components can be directly equated to the NP spin coefficients
of some normalized spinor dyad, were studied. It has been confirmed that such
Lanczos potentials exist in many special classes of spacetimes namely, many
stationary axially symmetric spacetimes and many cylindrically symmetric
spacetimes \cite{DM}, all conformally flat pure radiation spacetimes and all
Kerr-Schild spacetimes where $l^{a}$ is geodesic and shear-free \cite{AE3}.
Slight variations of the identification scheme also works for all type III,
N and 0 spacetimes \cite{ACL}. If we, in a class ${\cal G}$ spacetime, choose
a new normalized spinor dyad $(\xi_{0}^{A}, \xi_{1}^{A})$ from the spinor
fields in ${\cal S}^{A}$, then the components of the spinor $\Gamma_{ABCA'}$
are precisely the NP spin coefficients of the dyad $(\xi_{0}^{A},\xi_{1}^{A})$.
Hence, $L_{ABCA'}=\Gamma_{(ABC)A'}$ is a Lanczos potential of the Weyl spinor,
whose components can be directly equated to the spin coefficients in the manner
described in \cite{AE3}.

An important application of these results is the construction of quasi-local
momentum $P_{AA'}$ in spacetimes of class ${\cal G}$ given in the previous
section. The reason why we have not explored this application in greater detail
is that in order to examine the properties of $P_{AA'}$, and also of the
analogues of the Nester-Witten 2-form and the Sparling 3-form, in this more
general class of spacetimes, we would need to impose extra restrictions on the
global topology onto the class ${\cal G}$. Since we feel this would obscure the
results obtained so far, a detailed exploration of this application will be
postponed to a future paper. Another development of the Bergqvist-Ludvigsen 
connection in the Kerr spacetime is Harnett's \cite{Harnett} construction of
twistors for the Kerr spacetime. Hopefully the results in this paper could be
used to generalize this twistor construction to more general spacetimes.

\section*{Acknowledgements}

Special thanks are due to docent S. Brian Edgar for helpful 
suggestions and discussions.

\end{document}